\documentclass[fleqn,usenatbib]{mnras}

\usepackage{newtxtext,newtxmath}

\usepackage[T1]{fontenc}

\DeclareRobustCommand{\VAN}[3]{#2}
\let\VANthebibliography\thebibliography
\def\thebibliography{\DeclareRobustCommand{\VAN}[3]{##3}\VANthebibliography}

\usepackage{graphicx}	%
\graphicspath{{figures/}}
\usepackage{amsmath}	%
\usepackage{hyperref}
\usepackage[flushleft]{threeparttable}
\usepackage[scientific-notation=true]{siunitx}
\usepackage{physics}
\usepackage{siunitx}
\usepackage{xcolor}
\usepackage{subfig}
\usepackage{multirow}

\newcommand{\Msun}{\mathrm{M}_{\odot}}

\title[Cold vs. Hot Accretion and Angular Momentum in FIRE]{
Cold vs. Hot Gas Accretion and Angular Momentum in FIRE Simulations: From Halo to Galaxy Scales}

\author[I. Sultan et al.]{Imran Sultan,$^{1}$\thanks{E-mail: sultan@u.northwestern.edu}
Claude-André Faucher-Giguère,$^{1}$
Jonathan Stern,$^{2}$
Guochao Sun$^{1}$
\\
$^{1}$Center for Interdisciplinary Exploration and Research in Astrophysics (CIERA) and Department of Physics and Astronomy, Northwestern University,\\1800 Sherman Ave, Evanston, IL 60201, USA\\
$^{2}$School of Physics \& Astronomy, Tel Aviv University, Tel Aviv 69978, Israel\\
}

\date{Accepted XXX. Received YYY; in original form ZZZ}

\pubyear{2024}

\begin{document}
\label{firstpage}
\maketitle

\begin{abstract}
We present a systematic study of gas accretion and angular momentum in the circumgalactic medium (CGM) using high-resolution FIRE cosmological simulations. %
Our analysis includes halos crossing the critical $\sim10^{12}\ \Msun$ mass scale where several transitions 
have been found, including inner CGM virialization, the transition from bursty to steady star formation, and the emergence of thin disks.
We find the temperature of gas inflowing onto galaxies is correlated with the virialization of the inner CGM.
CGM inflows are almost entirely cold ($T<10^5$ K) in pre-virialized halos, while hot inflows ($T>10^5$ K) dominate in virialized halos. 
When hot inflows dominate, cooling generally occurs simultaneously with circularization at galaxy radii. 
The dominance of hot inflows onto massive galaxies persists even at high redshift where cold streams may coexist. %
Consistent with previous studies, cold inflows have higher specific angular momentum than dark matter and hot gas. %
Yet, we find in bursty, low-mass galaxies, cold inflows do not circularize prior to star formation, while in steady, massive galaxies, hot inflows circularize, cool, and form stars with disk-like kinematics. 
We additionally find that in bursty galaxies, accreted gas typically forms stars after residing in the galaxy for less than $\sim5$ galaxy free-fall times, while in steady galaxies, gas can persist in the galaxy for up to $\sim25$ free-fall times before forming stars.
This highlights a key difference between star formation in bursty galaxies fed by cold accretion and steady equilibrium disks fed by hot accretion. 

\end{abstract}


\section{Introduction}\label{sec:Intro}
A growing number of studies suggest the role of the circumgalactic medium (CGM) in galaxy formation may be significant, although it is not yet fully understood.
The CGM, the gas contained inside the dark matter halo (extending to $\sim R_{\mathrm{vir}}$, although there is no strict boundary), and outside the galaxy's interstellar medium (ISM), is the intermediary between the intergalactic medium (IGM) and the galaxy. 
All of the gas that fuels star formation and active galactic nuclei in the galaxy first flows through the CGM, and thus understanding the physics of gas flows in the CGM may be essential for understanding the evolution of the central galaxy.
For reviews of observational and theoretical work in the field, see reviews by, e.g., \cite{tumlinsonCircumgalacticMedium2017} and \cite{faucher-giguereKeyPhysicalProcesses2023}.
Key questions remain regarding how the CGM couples to, e.g., the formation of rotationally-supported disks, the transition from bursty to steady star formation, and the quenching of star formation.

There has been a large set of literature exploring gas accretion through the CGM, including how mass and angular momentum are delivered onto galaxies in a broad set of simulations.
In the classical picture, gas with a cooling time ($t_{\mathrm{cool}}$: the time to radiate away all internal energy) longer than the free-fall time ($t_{\mathrm{ff}}$: the time for gravitational collapse to the center of the gravitational potential) is able to maintain a hot virialized steady state at temperatures $T \sim T_{\mathrm{vir}}$.
In this picture, gaseous halos virialize when gravitationally collapsing gas enters the CGM from the IGM at supersonic speeds, creating a shock that heats the gas to temperatures similar to the virial temperature of the halo.
On the other hand, gas with a cooling time shorter than the free-fall time quickly cools and accretes onto the galaxy in cold clouds or streams.
The halo mass threshold that marks the transition from cold to hot accretion is $M_{\mathrm{halo}} \sim 10^{11} -10^{12}\ \Msun$, with the exact value depending on CGM mass, metallicity, radius, and redshift (e.g., \citealt{reesCoolingDynamicsFragmentation1977, whiteCoreCondensationHeavy1978, birnboimVirialShocksGalactic2003, keresHowGalaxiesGet2005, keresGalaxiesSimulatedLCDM2009, faucher-giguereSmallCoveringFactor2011, sternMaximumAccretionRate2020, goldnerAccretiondrivenTurbulenceCircumgalactic2026}).

Although classical theory predicts that virialized halos of mass $\sim 10^{11} -10^{12}\ \Msun$ are almost completely hot at $z \sim 0$, at higher redshifts $z \gtrsim 1.5$, the CGM is predicted by simulations and analytical results to be multiphase, with significant cold streams penetrating the hot halo (e.g., \citealt{keresHowGalaxiesGet2005, dekelGalaxyBimodalityDue2006, dekelColdStreamsEarly2009, dekelFormationMassiveGalaxies2009, vandevoortRatesModesGas2011, medlockStatisticalPropertiesCold2025}).
One reason for this, suggested by \cite{dekelGalaxyBimodalityDue2006}, is related to the different cosmological environments of $\sim 10^{11} -10^{12}\ \Msun$ halos at high and low redshifts.
Halos of mass $\sim 10^{11} -10^{12}\ \Msun$ are increasingly rare at higher redshifts, and thus are found in the nodes of the cosmic web, where they are fed by a few filaments with cross sections smaller than the halos.
Cold streams originate from the filaments, which have high densities and therefore cool efficiently and remain cold as they flow to the center of the halo.
In contrast, at $z \sim 0$, halos at this mass range are more common and may be found embedded in much larger filaments.
\cite{dekelGalaxyBimodalityDue2006} derived a rough analytical model for the maximum halo mass of virialized halos at $z \gtrsim 1.5$ that can contain cold streams, $M_{\mathrm{stream}}$ (see also \citealt{waterval_gas_2025} for a recent prediction of $M_{\mathrm{stream}}$ using cosmological zoom-in simulations).
Some work in this field suggests that the cold streams may be the primary mode of gas delivery to galaxies (e.g., \citealt{dekelColdStreamsEarly2009}).

The angular momentum of the circumgalactic medium, and how it correlates with that of the dark matter halo and the central galaxy, may also play a key role in disk formation but is not well understood (for a review, see, e.g., \citealt{faucher-giguereKeyPhysicalProcesses2023}).
Galaxies and their CGM acquire angular momentum from their host dark matter halo in the standard cosmological theory (e.g., \citealt{fallFormationRotationDisc1980}), although hydrodynamic forces and feedback from galaxies can then significantly modify the angular momentum of the baryons.
Studies of simulations have found systematic differences in the specific angular momentum (sAM) of the dark matter halo and CGM, and between different CGM phases.
In particular, the CGM is found to have a higher amount of sAM than its dark matter halo, including in simulations which neglected radiative cooling \citep{zjupaAngularMomentumProperties2017}.
A variety of simulations have shown that the dominant source of this difference is cold CGM gas, which contains significantly more sAM than gas in the hot phase (e.g., \citealt{danovichFourPhasesAngularmomentum2015, stewartHighAngularMomentum2017, defelippisAngularMomentumCircumgalactic2020, wangLargescaleEnvironmentCGM2021}).

A possible reason for this trend in sAM is the thin geometry of cold streams, which allows it to be more effectively gravitationally torqued before entering the halo, in contrast to the more isotropic dark matter distribution \citep{danovichFourPhasesAngularmomentum2015}.
Other causes are that cold gas, which can be accreted on much shorter timescales than hot gas, may serve as a tracer of the large-scale environment, which has more sAM (e.g., \citealt{stewartANGULARMOMENTUMACQUISITION2013}), and that galaxy feedback may preferentially eject gas with low sAM from the halo (e.g., \citealt{brookHierarchicalFormationBulgeless2011, agertzImpactStellarFeedback2016}).

Recent studies of the Feedback in Realistic Environments (FIRE) simulations have shed new light on the role of the CGM in galaxy formation.
\cite{sternVirializationInnerCGM2021} found that halos virialize from the outside in, with the inner CGM virialzing when the outer CGM is already at the virial temperature.
They analyzed FIRE-2 simulations over a broad range of redshifts and halo masses, and found that the virialization of the inner CGM (inner-CGM virialization, ICV), which happens when the cooling time of the hot gas exceeds roughly a free-fall time at $\sim 0.1 R_{\mathrm{vir}}$, occurs when the halo reaches a mass of $\sim 10^{12}\ \Msun$, with no significant dependence on redshift.
Additional studies have correlated the virialization of the inner CGM with several key transitions in galaxy properties, including the transition from dispersion-dominated kinematics to the formation of large, thin galactic disks \citep{hafenHotmodeAccretionPhysics2022, sternVirializationInnerCGM2021, myrtajProtogalaxyThickThin2026}, highly time variable (`bursty') to more steady star formation rates, a reduction in stellar feedback-driven gas outflows from the ISM \citep{muratovGustyGaseousFlows2015}, and inefficient to accelerated supermassive black hole (SMBH) growth \citep{byrneStellarFeedbackregulatedBlack2023}.
\cite{kakolyTurbulencedominatedCGMOrigin2025} and \cite{goldnerAccretiondrivenTurbulenceCircumgalactic2026} also found that ICV coincides with a transition in the inner halo pressure support from being turbulence-dominated to thermal-dominated.

In particular, \cite{hafenHotmodeAccretionPhysics2022} studied FIRE zoom-in simulations of Milky Way-mass halos at $z\sim0$, and linked the emergence of thin galaxy disks to accretion in the inner CGM dominated by hot, rotating cooling flows (see also \citealt{sternAccretionDiscGalaxies2024, sankarHotAccretionSpiral2025}).
Once the inner CGM virializes, gas with cooling times greater than the free-fall time but shorter than the Hubble time can form such a `cooling flow,' in which it flows inward to the potential center while radiatively cooling \citep{sternCoolingFlowSolutions2019, sternMaximumAccretionRate2020}.
\cite{sultanCoolingFlowsReference2025} showed that CGM-scale analytic cooling flow models reproduce the thermodynamic properties of the hot CGM in $\sim10^{12}$–$10^{13}\ \Msun$ FIRE halos without AGN feedback, and that cooling flows are consistent with a range of low-redshift observations of $\sim 10^{12}\ \Msun$ halos.
In contrast, for $\sim 10^{13}\ \Msun$ halos, \cite{sultanCoolingFlowsReference2025} found cooling flows predict X-ray surface brightness profiles from thermal emission that are much steeper than recent stacked observations, likely because the idealized models neglect AGN feedback.
\cite{mccarthyCaseAGNFeedback2010} found that although simulations with feedback only from stars can reproduce some observed thermodynamic properties of the hot gas in group-mass halos, AGN feedback is needed to produce realistic central galaxies with stellar massses and star formation rates consistent with observations, and to prevent the overly massive galaxies that form in models with only stellar feedback (see also \citealt{wellonsExploringSupermassiveBlack2023} and \citealt{byrneEffectsMultichannelActive2024} for recent FIRE studies which similarly found that including AGN feedback produces galaxies with more realistic properties in $\sim 10^{13}\ \Msun$ halos).

The finding that thin MW-mass disks at $z\sim0$ are dominantly fed by hot rotating inflows in FIRE contrasts with a common picture of galaxy formation in which cold streams are the primary mode of gas delivery to galaxies, and promote disk formation by coherently delivering high amounts of specific angular momentum (e.g., \citealt{dekelColdStreamsEarly2009}).
\cite{{yuBornThisWay2023}} similarly showed that thin-disk stellar populations in FIRE form after the bursty phase of star formation and the virialization of the inner CGM.

Motivated by these recent developments, in this paper we perform a systematic analysis of accretion and angular momentum transport in FIRE simulations over a wide mass and redshift range.
We study FIRE halos with mass spanning $M_{\mathrm{halo}} \sim 10^{10.5}-10^{13}\ \Msun$, over redshifts $0 \lesssim z \lesssim 5$.
We investigate the accretion of hot and cold gas, and quantify the delivery of angular momentum by the two accretion modes over a wide range of physical scales, from the scale of the halo ($\sim R_\mathrm{vir}$) down to the scale of the galaxy ($\lesssim 0.05R_\mathrm{vir}$).
Our sample includes high-redshift halos predicted to be in the `cold-in-hot' regime, with cold streams penetrating the hot, virialized halo. 
The primary questions we explore are what the primary mode is by which gas and angular momentum are delivered to galaxies, whether inflows circularize, and what the implications of these are for star formation and the formation of galactic disks.

We analyze cosmological zoom-in simulations carried out by the FIRE project\footnote{\url{https://fire.northwestern.edu}}.
Cosmological zoom-ins combine the large-scale dark matter information with high-resolution hydrodynamical simulations focused on selected galaxies, enabling the modeling of galaxies and their CGM in their cosmological context.
As a result of the detailed models of star formation and stellar feedback, FIRE simulations reproduce a range of observed galaxy properties up to Milky Way masses, including stellar masses \citep{hopkinsGalaxiesFIREFeedback2014, hopkinsFIRE2SimulationsPhysics2018, feldmannFIREboxSimulatingGalaxies2023}, mass-metallicity relations (e.g., \citealt{maOriginEvolutionGalaxy2016, bassiniInflowOutflowProperties2024, marszewskiHighRedshiftGasPhaseMass2024}), and galaxy structural properties (e.g., \citealt{el-badryGasKinematicsMorphology2018}).

\begin{table*}
\begin{center}
\caption{FIRE simulations used in this work. 
For each halo, we list the halo virial mass, stellar mass, and halo virial radius
at select integer redshifts, as well as the baryonic mass resolution
($m_b$), the redshift at which the inner CGM is expected to virialize
($z_{\mathrm{ICV}}$), and the redshift at which the central galaxy transitions from bursty (i.e., highly time variable) to steady star formation ($z_{\mathrm{bursty}}$).
The halos are sorted in ascending order by $M_\mathrm{vir}(z=1)$.
}
\begin{tabular}{c|c|c|c|c|c|c|c}
\hline
Halo Name &
$z$ &
$M_{\mathrm{vir}}$ [$\Msun$] &
$M_*$ [$\Msun$] &
$R_{\mathrm{vir}}$ [pkpc] &
$m_b$ [$\Msun$] &
$z_{\mathrm{ICV}}$ &
$z_{\mathrm{bursty}}$ \\
\hline

\multirow{4}{*}{m12b} &
3 &
\num[round-mode=places,round-precision=1]{117169658119.65813} &
\num[round-mode=places,round-precision=1]{374000780.28440475} &
40 &
\multirow{4}{*}{\num[round-mode=places,round-precision=0]{7000.0}} &
\multirow{4}{*}{0.70} &
\multirow{4}{*}{0.76} \\
&
2 &
\num[round-mode=places,round-precision=1]{458015669515.66956} &
\num[round-mode=places,round-precision=1]{3794276118.2785034} &
84 & & & \\
&
1 &
\num[round-mode=places,round-precision=1]{691675213675.2137} &
\num[round-mode=places,round-precision=1]{29396307468.414307} &
138 & & & \\
&
0 &
\num[round-mode=places,round-precision=1]{1305954415954.416} &
\num[round-mode=places,round-precision=1]{90475053787.23145} &
286 & & & \\
\hline

\multirow{4}{*}{m12f} &
3 &
\num[round-mode=places,round-precision=1]{170038461538.46155} &
\num[round-mode=places,round-precision=1]{446125343.4419632} &
45 &
\multirow{4}{*}{\num[round-mode=places,round-precision=0]{7000.0}} &
\multirow{4}{*}{0.44} &
\multirow{4}{*}{0.55} \\
&
2 &
\num[round-mode=places,round-precision=1]{470625356125.35614} &
\num[round-mode=places,round-precision=1]{3158679604.5303345} &
84 & & & \\
&
1 &
\num[round-mode=places,round-precision=1]{722783475783.4758} &
\num[round-mode=places,round-precision=1]{14720352888.1073} &
141 & & & \\
&
0 &
\num[round-mode=places,round-precision=1]{1490099715099.715} &
\num[round-mode=places,round-precision=1]{55157051086.42578} &
299 & & & \\
\hline

\multirow{4}{*}{m12i} &
3 &
\num[round-mode=places,round-precision=1]{153545584045.58405} &
\num[round-mode=places,round-precision=1]{137905748.56102467} &
44 &
\multirow{4}{*}{\num[round-mode=places,round-precision=0]{7000.0}} &
\multirow{4}{*}{0.56} &
\multirow{4}{*}{0.56} \\
&
2 &
\num[round-mode=places,round-precision=1]{287898860398.8604} &
\num[round-mode=places,round-precision=1]{868354216.2179947} &
72 & & & \\
&
1 &
\num[round-mode=places,round-precision=1]{843410256410.2565} &
\num[round-mode=places,round-precision=1]{14664174318.313599} &
149 & & & \\
&
0 &
\num[round-mode=places,round-precision=1]{1101970085470.0854} &
\num[round-mode=places,round-precision=1]{55143885612.48779} &
270 & & & \\
\hline \hline

\multirow{3}{*}{m13A1} &
3 &
\num[round-mode=places,round-precision=1]{1689540889526.5425} &
\num[round-mode=places,round-precision=1]{91263074874.87793} &
97 &
\multirow{3}{*}{\num[round-mode=places,round-precision=0]{30000.0}} &
\multirow{3}{*}{3.6} &
\multirow{3}{*}{3.4} \\
&
2 &
\num[round-mode=places,round-precision=1]{2844461979913.917} &
\num[round-mode=places,round-precision=1]{172087688446.04492} &
153 & & & \\
&
1 &
\num[round-mode=places,round-precision=1]{3927905308464.8496} &
\num[round-mode=places,round-precision=1]{272157363891.60156} &
247 & & & \\
\hline

\multirow{3}{*}{m13A4} &
3 &
\num[round-mode=places,round-precision=1]{941430416068.8666} &
\num[round-mode=places,round-precision=1]{14562433958.053589} &
80 &
\multirow{3}{*}{\num[round-mode=places,round-precision=0]{30000.0}} &
\multirow{3}{*}{2.2} &
\multirow{3}{*}{2.0} \\
&
2 &
\num[round-mode=places,round-precision=1]{3059913916786.227} &
\num[round-mode=places,round-precision=1]{119040250778.19824} &
156 & & & \\
&
1 &
\num[round-mode=places,round-precision=1]{4569856527977.045} &
\num[round-mode=places,round-precision=1]{225249595642.08984} &
260 & & & \\
\hline

\multirow{3}{*}{m13A2} &
3 &
\num[round-mode=places,round-precision=1]{2043529411764.706} &
\num[round-mode=places,round-precision=1]{117964792251.58691} &
103 &
\multirow{3}{*}{\num[round-mode=places,round-precision=0]{30000.0}} &
\multirow{3}{*}{2.7} &
\multirow{3}{*}{2.9} \\
&
2 &
\num[round-mode=places,round-precision=1]{3667718794835.0073} &
\num[round-mode=places,round-precision=1]{282737388610.83984} &
166 & & & \\
&
1 &
\num[round-mode=places,round-precision=1]{7753142037302.727} &
\num[round-mode=places,round-precision=1]{405769119262.6953} &
310 & & & \\
\hline

\multirow{3}{*}{m13A8} &
3 &
\num[round-mode=places,round-precision=1]{803301291248.2067} &
\num[round-mode=places,round-precision=1]{5951787829.399109} &
76 &
\multirow{3}{*}{\num[round-mode=places,round-precision=0]{30000.0}} &
\multirow{3}{*}{1.9} &
\multirow{3}{*}{2.3} \\
&
2 &
\num[round-mode=places,round-precision=1]{1514304160688.6658} &
\num[round-mode=places,round-precision=1]{64752998352.05078} &
124 & & & \\
&
1 &
\num[round-mode=places,round-precision=1]{12623486370157.82} &
\num[round-mode=places,round-precision=1]{488695449829.10156} &
359 & & & \\
\hline

\end{tabular}\label{tab:sims}
\end{center}
\end{table*}

This paper is organized as follows.
We begin in Section \ref{sec:Methods} with a description of the FIRE simulations we analyze and our halo analysis methods, including our criterion for finding the time of virialization of the inner halo.
In the following three sections, we present our results over three physical scales:
In Section \ref{sec:ResultsI}, we show results of the accretion rate and angular momentum of hot and cold gas on halo scales.
In Section \ref{sec:ResultsII}, we track particles inflowing from the inner CGM to the central galaxy, and show results on the physics of the accretion.
After the particles are accreted onto the central galaxy, we continue tracking the gas (as it undergoes star formation or ejection), and show results of its properties in Section \ref{sec:ResultsIII}.
Finally, we provide a discussion of our results in Section \ref{sec:Discussion} and summarize the main findings of this study in Section \ref{sec:Conclusions}.

Throughout this paper we use $\log$ to refer to the base-10 logarithm.
We use a standard flat $\Lambda$CDM cosmology consistent with $h \approx 0.67$, $\Omega_{m} \approx 0.3$, and  $\Omega_{b} \approx 0.05$ \citep{aghanimPlanck2018Results2020}.

\section{Simulations and Methods}\label{sec:Methods}
In this section we describe the FIRE simulations used in this work, as well as our halo analysis methods.

\subsection{FIRE Simulations}\label{sec:FIREmethod}
The cosmological zoom-in simulations used in this study are from the FIRE project.
The simulations were run with the GIZMO\footnote{\url{http://www.tapir.caltech.edu/~phopkins/Site/GIZMO.html}} gravity+hydrodynamics code \cite{hopkinsNewClassAccurate2015} using a meshless finite-mass hydrodynamics method.
Considerable progress has been made by the FIRE project in modeling the physics of stellar feedback and the multi-phase interstellar medium (ISM). 
For example, the rates of Type-Ia and Type-II supernovae explosions are modeled separately, and feedback including mass, metals, energy, and momentum are injected into the nearby ISM. 
Star particles also lose mass by stellar winds for both OB and AGB stars.
Radiative feedback models that take into account multiple wavelengths are utilized to simulate the photoionization and photoelectric heating effects, and the corresponding radiation pressure is also modeled.
A complete description of the FIRE-2 methods is given in \cite{hopkinsFIRE2SimulationsPhysics2018}. 

Our analysis includes halos that span a wide range in mass, $M_{\mathrm{halo}} \sim 10^{10.5}-10^{13}\ \Msun$, over redshifts $0 \lesssim z \lesssim 5$.
The FIRE-2 simulation set we analyze in this work is given by Table \ref{tab:sims}, and consists of three low-redshift Milky Way-mass halos (our `m12' halos) and four high-redshift more massive halos (our `m13' halos), where each halo represents a different initial condition for the zoom-in simulation.

The three low-redshift MW-mass halos in our set, which reach halo masses $\sim 10^{12}\ \Msun$ by $z =0$, are part of the `core' FIRE-2 simulation suite (see \citealt{wetzelPublicDataRelease2023, wetzelSecondPublicData2025} for the FIRE-2 public data release, and additionally \citealt{hopkinsFIRE2SimulationsPhysics2018}).
The simulations were run with the default FIRE physics models, i.e., simulations with no black holes and no cosmic rays and evolved to $z=0$.
These simulations include a subgrid model for the turbulent diffusion of metals in gas \citep{colbrookScalingLawsPassivescalar2017, escalaModellingChemicalAbundance2018}.

In addition, we analyze FIRE-2 simulations of more massive high-redshift halos carried out by \cite{angles-alcazarBlackHolesFIRE2017}, which reach halo masses of $\sim 10^{12.6} - 10^{13}\ \Msun$ by $z=1$.
These four halos were first presented in \cite{feldmannFormationMassiveQuiescent2016, feldmannColoursStarFormation2017a},
where they were evolved with an earlier version of the FIRE code \citep{hopkinsGalaxiesFIREFeedback2014}.
The FIRE-2 runs from \cite{angles-alcazarBlackHolesFIRE2017} include black hole growth but neglect AGN feedback.
We do not expect the inclusion of black hole accretion to play a significant role in the halo and galaxy scale results we present in this work.
However, the exclusion of black hole feedback in the massive halo simulations we analyze resulted in overly compact stellar cores and excessive star formation in the massive galaxies at late times, which is why the simulations were run to $z=1$ \citep{wellonsMeasuringDynamicalMasses2020, parsotanRealisticMockObservations2021, byrneStellarFeedbackregulatedBlack2023}.
Although AGN feedback is likely needed to quench star formation and prevent overly compact stellar cores in massive halos (e.g., \citealt{byrneEffectsMultichannelActive2024}), the simulations we analyze in this work are a useful tool to study gas accretion in the absence of complicating effects such as AGN feedback.

In Table \ref{tab:sims}, we list properties of the halos and their central galaxies at integer redshifts from $z=3$ to the final redshift simulated.
The initial mass of gas and star particles (i.e. the baryonic mass resolution) is $\num{7e3}\ \Msun$ for the m12 simulations and $\num{3e4}\ \Msun$ for the m13 simulations.
Dark matter particles have a mass that is a factor of $(\Omega_m - \Omega_b)/\Omega_b \approx 5$ larger than the baryonic resolution elements in the high-resolution region.

For gas resolution elements, gravitational softening is treated in an adaptive manner, where it is set equal to the smoothing length of the gas.
The minimum (Plummer equivalent) force softening length for gas ranges from 0.4 pc to 1 pc for the m12 halos, and is 0.7 pc for the m13 halos (physical units).
The gravitational softening is fixed for star and dark matter particles at $z<9$.
The Plummer equivalent force softening length for stars is 4 pc (7 pc), and for dark matter is 40 pc (57 pc) for the m12 (m13) halos, in physical units.

\subsection{Galaxy centering}\label{sec:methods_halocentering}
To find the center of the central galaxy in each snapshot, we first find the center coordinates of the host halo (defined as the center of mass of dark+baryonic matter in the halo), as well as its virial radius and virial mass, using the Amiga Halo Finder \citep{knollmannAhfAmigasHalo2009}.
Consistent with \cite{bryanStatisticalPropertiesXRay1998}, we define the virial radius $R_{\mathrm{vir}}$ of the halo as the radius of a sphere centered on the halo with density $\Delta_{\mathrm{vir}} \rho_{\mathrm{crit}}$, where $\rho_{\mathrm{crit}}(z)$ is the critical density of the universe at redshift $z$ and we use their fitting function for the virial overdensity: $\Delta_{\mathrm{vir}}(z) = 18 \pi^2 +82 x - 39 x^2,$ where
$x(z) = \Omega_m(z) - 1$.
The virial mass $M_{\mathrm{vir}}$ is the total mass enclosed within the sphere.

Since the galaxy can be offset from the center of mass of the host halo, we determine the galaxy center using the iterative center of mass method of \cite{powerInnerStructureLCDM2003}.
We start by finding the center of mass of all star particles within $0.2 R_{\mathrm{vir}}$ of the halo center, and iteratively recompute the center of mass within a shrinking aperture until the enclosed stellar mass is $\sim 1\%$ of the mass in the initial aperture. 
We verified that our results in this work are insensitive to the exact aperture choices by varying the initial and final apertures.

At each snapshot, we rotate particle coordinates and velocities such that the z-axis is the axis of rotation of the central galaxy.
Following \cite{gurvich_rapid_2023}, we define the galaxy's rotation axis as the total angular momentum of all star particles within $5 R_{s,1/2}$, where $R_{s,1/2}$ is the radius that encloses 50\% of the total stellar mass contained within $0.2 R_{\mathrm{vir}}$.
We measure particle velocities in the rest-frame of the galaxy by subtracting the velocity of the center of mass of stars within $5 R_{s,1/2}$ of the galaxy center.
Finally, we define the stellar mass and gas mass of the galaxy, $M_*$ and $M_{\mathrm{gas}}$, as the total mass of all stellar and gas particles within $0.05R_{\mathrm{vir}}$ of the center of the halo.
In the results we present in this work, we commonly use $0.05 R_{\mathrm{vir}}$ as an approximation of the galaxy radius rather than $5 R_{s,1/2}$, as the former tends to more smoothly evolve with time compared to estimates based on $R_{s,1/2}$, which are more prone to fluctuations at small time scales, especially in low-mass galaxies undergoing bursts of star formation and resulting feedback.

\begin{figure*}
    \subfloat{\includegraphics[width=6.5in]{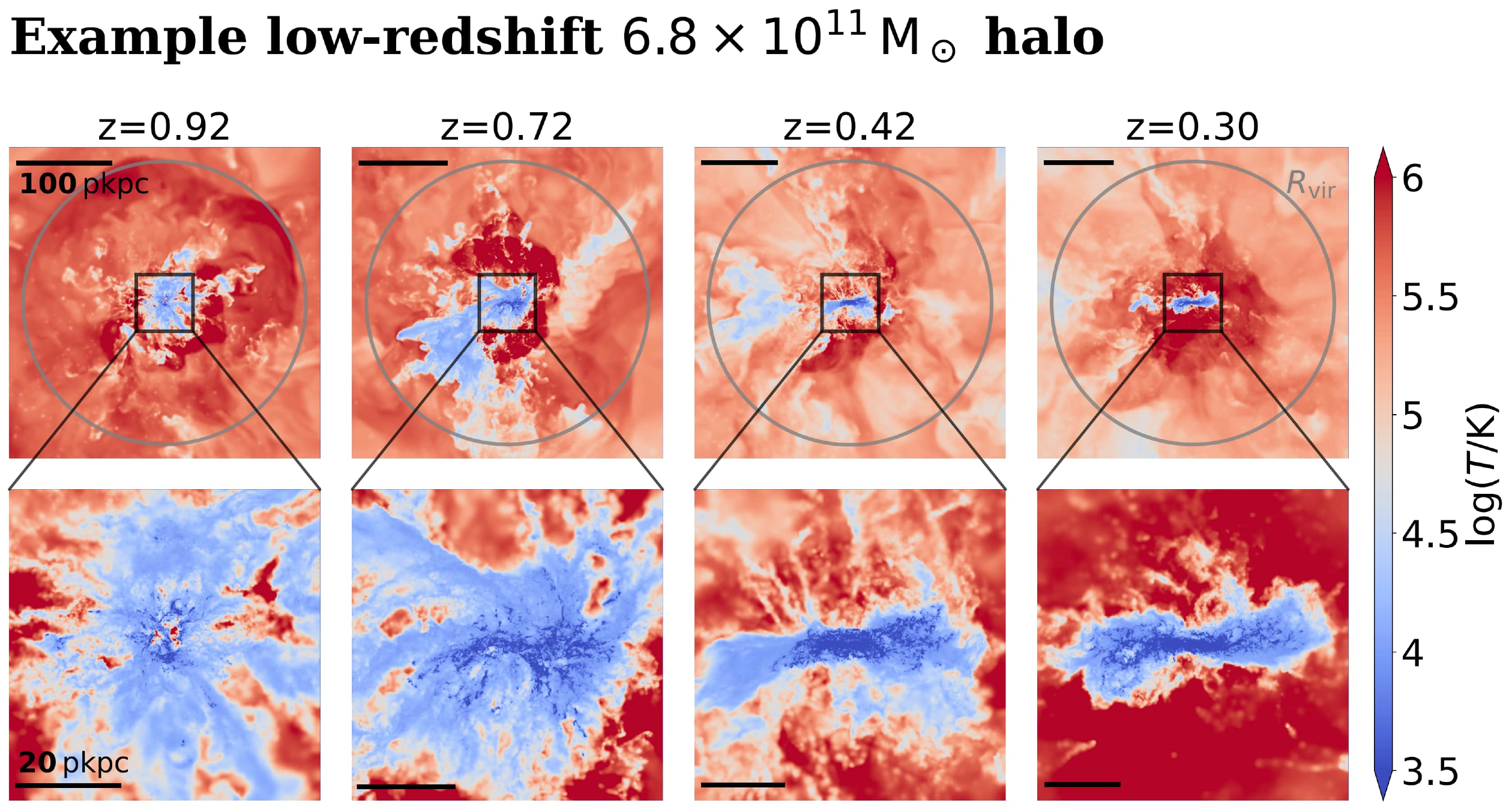}}
    \\[3em]
     \subfloat{\includegraphics[width=6.5in]{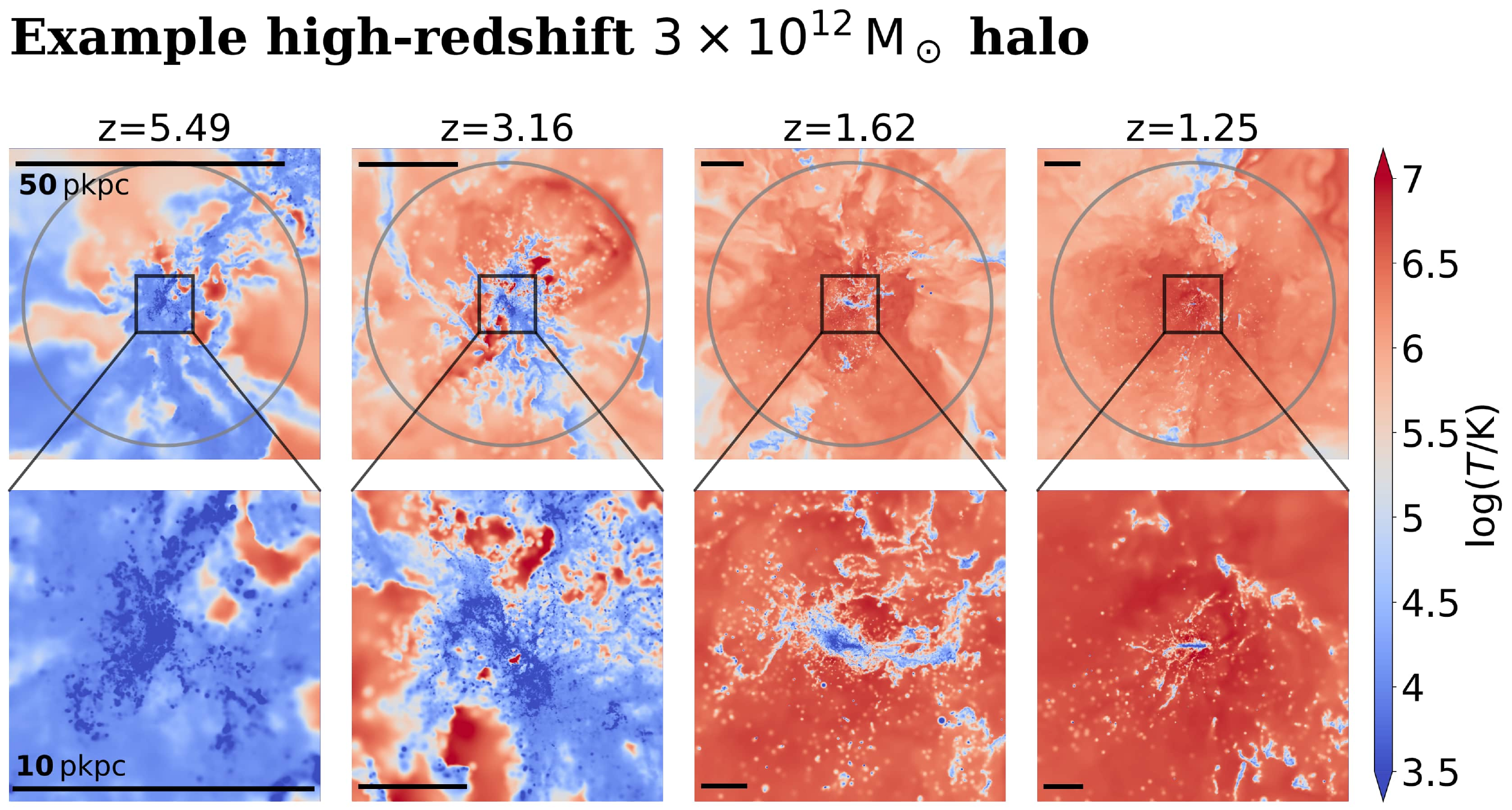}}
      \caption{Gas temperature maps for two representative FIRE halos in our analysis set: m12i (top) and m13A4 (bottom).
    Projections of $\log T$ weighted by mass through a 30 comoving kpc slice perpendicular to the galactic plane are shown, giving an edge-on view of each galaxy.
    The galactic plane is defined as the plane orthogonal to the rotation axis of the galaxy, which we compute from the total angular momentum of all star particles within $5 R_{s,1/2}$ (see Section \ref{sec:methods_halocentering}). 
    Temperature maps at four simulation snapshots are shown for each halo, corresponding to times $t - t_{\mathrm{ICV}} = \{-2, -1, 1, 2\}$ Gyr.
    The masses of the halos in the last snapshots shown are given.
    The upper panel spans $2.2 R_{\mathrm{vir}}$, while the lower panel shows a zoomed-in view of the inner CGM with side length $0.4 R_{\mathrm{vir}}$.
    In each row, the scale bar indicates the length value labeled in the first column, in units proper kpc.
    The halos initially contain significant amounts of cold gas in their inner regions ($\sim 0.1 R_{\mathrm{vir}}$).
    The inner halos become increasingly hot gas-dominated as the simulations are evolved.
    }
    \label{fig:Tmaps_twoscales}
\end{figure*}

\subsection{Inner CGM Virialization}\label{sec:tICV}

Throughout our paper, $z_{\mathrm{ICV}}$ and $t_{\mathrm{ICV}}$ refer to the approximate redshift and corresponding cosmic time at which the inner CGM virializes (ICV).
Redshifts $z_{\mathrm{ICV}}$ for the halos in our analysis set are given in Table $\ref{tab:sims}$, which we estimate as described below.
We use the redshift of ICV as a rough indicator of the redshift when the inner halo for a given galaxy transitions from being dominated by cold gas to hot gas. 

Following \cite{sternVirializationInnerCGM2021}, we use the ratio of the estimated cooling time of shock-heated virial-temperature gas in the inner CGM to the free-fall time, $t_{\mathrm{cool}}^{(s)}/t_{\mathrm{ff}}$, to approximate when the inner CGM virializes in our halos.
$t_{\mathrm{cool}}^{(s)}$ and $t_{\mathrm{ff}}$ are calculated at $0.1R_{\mathrm{vir}}$ as described in their Section 2.4.
Briefly, the free-fall time is given by 
\begin{equation}
    t_{\mathrm{ff}}(r) = \frac{\sqrt{2}r}{v_c(r)},
\end{equation}
where the circular velocity $v_c \equiv \sqrt{\frac{G M(<r)}{r}}$ is the velocity in the gravitational potential at a radial distance $r$, and $M(<r)$ is the total mass contained within a distance $r$ from the halo center.
This expression is the time for a particle at radius $r$ to radially infall to the center under constant gravitational acceleration; we adopt this definition for consistency with previous CGM studies (e.g., \citealt{sternVirializationInnerCGM2021, mccourtThermalInstabilityGravitationally2012}).
$t_{\mathrm{cool}}^{(s)}$ is the cooling time of virial-temperature shocked gas, defined as
\begin{equation}
    t_{\mathrm{cool}}^{(s)} \equiv t_{\mathrm{cool}}(T^{(s)}, n_{\mathrm{H}}^{(s)})=\frac{3/2 \cdot 2.3 k_\mathrm{B} T^{(s)}}{n_{\mathrm{H}}^{(s)} \Lambda},
\end{equation}
where $\Lambda$ is the cooling function and we assume fully ionized gas with $n/n_\mathrm{H}=2.3$. 
We use $T^{(s)}$ and $n_{\mathrm{H}}^{(s)}$, which are estimates of the temperature and density of virialized gas in the inner CGM in hydrostatic equilibrium ($s$ for `shocked'), rather than $T$ and $n_{\rm H}$ in the simulation, so the cooling time calculation will be valid even pre-ICV when most of the gas has a temperature $\ll T_{\rm vir}$. 
Specifically, following \cite{sternVirializationInnerCGM2021} we use
\begin{align}
    T^{(s)} &= \frac{2 \mu m_p v_c^2}{3k_\mathrm{B}}\\
    n_{\mathrm{H}}^{(s)} &= \frac{1}{2.3 k_\mathrm{B} T^{(s)}} \int_{0.1R_\mathrm{vir}}^{R_\mathrm{vir}}\frac{\rho v_c^2}{r} \dd{r},
\end{align}
where $m_p$ is the mass of a proton and we assume a mean molecular weight $\mu=0.62$ for a fully ionized plasma of primordial composition.
The expression for $n_{\mathrm{H}}^{(s)}$ follows from integrating the equation of hydrostatic equilibrium and assuming that the pressure at $R_\mathrm{vir}$ is negligible compared to the pressure at $0.1R_\mathrm{vir}$.
The integral in $n_{\mathrm{H}}^{(s)}$ represents the weight of the overlying gas in the halo.
The gas density $\rho$ is calculated as the ratio of the total gas mass to the total volume in radial shells with a width of 0.05 dex in $\log r$. 
Using the mean density in the shell rather than $n_{\rm H}^{(s)}$ has a minor effect on $t_{\rm cool}^{(s)}$ (see \citealt{sternVirializationInnerCGM2021}). 

\cite{sternVirializationInnerCGM2021} analyzed FIRE-2 simulations over a wide range of halo masses and redshifts, and found that the inner CGM virializes, i.e. becomes primarily hot, when $1 < t_{\mathrm{cool}}^{(s)}/t_{\mathrm{ff}} <4$.
Following their approach, we adopt the intermediate value $t_{\mathrm{cool}}^{(s)}/t_{\mathrm{ff}} = 2.5$ to approximate when the inner CGM virializes in our halos.

An exception is m12i.
As shown in Figure \ref{fig:tbursty_tICV}, m12i has an unusually flat evolution of $t_{\mathrm{cool}}^{(s)}/t_{\mathrm{ff}}$ near 2.5 (in contrast to the other halos), and the exact time the ratio crosses this fiducial value is poorly defined,
For m12i, we therefore set $z_{\mathrm{ICV}}$ to the redshift associated with the transition in the galaxy from bursty to steady star formation (see Section \ref{sec:burstytosteadySFR}), motivated by previous FIRE studies suggesting that the star formation rate stabilizes near ICV (e.g., \citealt{sternVirializationInnerCGM2021, gurvich_rapid_2023}).
For all of the other halos in our sample, this redshift has close agreement with the estimate for $z_{\mathrm{ICV}}$ using the $t_{\mathrm{cool}}^{(s)}/t_{\mathrm{ff}}=2.5$ threshold (see Figure \ref{fig:tbursty_tICV}).

In Figure \ref{fig:Tmaps_twoscales}, we show maps of temperature for two example halos in our analysis, with an edge-on view of the galactic plane.
The visualizations were produced with FIRE Studio \citep{gurvichFIREStudioMovie2022}, and show a 30 comoving kpc slice (i.e., a projection of all gas mass within $|x|<15$ comoving kpc on the yz-plane), where the halo has been rotated such that the total angular momentum vector of star particles in the galaxy lies along the z-axis.
The temperature maps at four snapshots are shown, spanning times 2 Gyr before ICV to 2 Gyr after ICV.
At each time, we show a halo-scale panel (with a side length of $2.2 R_{\mathrm{vir}}$), and a zoomed-in view of the inner halo ($0.4 R_{\mathrm{vir}}$).
Generally, the halos initially contain significant amounts of cold gas, including in their inner regions near $\sim 0.1 R_{\mathrm{vir}}$.
The halos become increasingly dominated by hot gas as time progresses to ICV and beyond.

At the last snapshot shown, corresponding to 2 Gyr after ICV, the galaxies are enveloped by a hot medium of $\sim T_{\mathrm{vir}}$.
The formation of galactic disks is apparent in the edge-on projections, which reveal a cold ISM.
The central disks that form in the high redshift massive halos are extremely compact (see Section \ref{sec:FIREmethod}).

\begin{figure*}
    \subfloat{\includegraphics[width=0.4\textwidth]{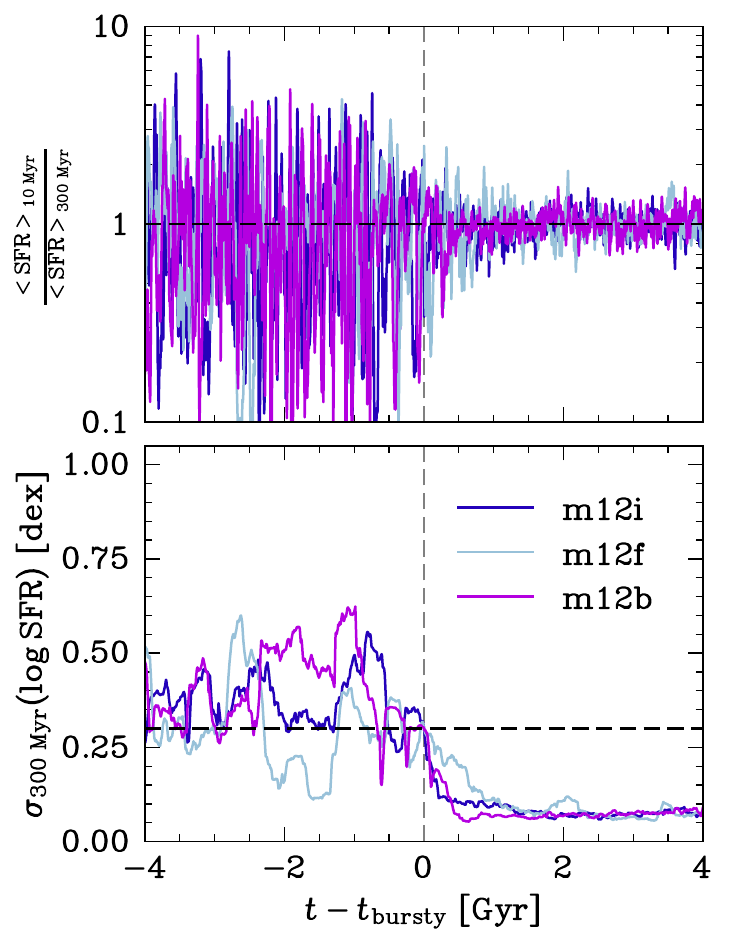}}
    \subfloat{\includegraphics[width=0.4\textwidth]{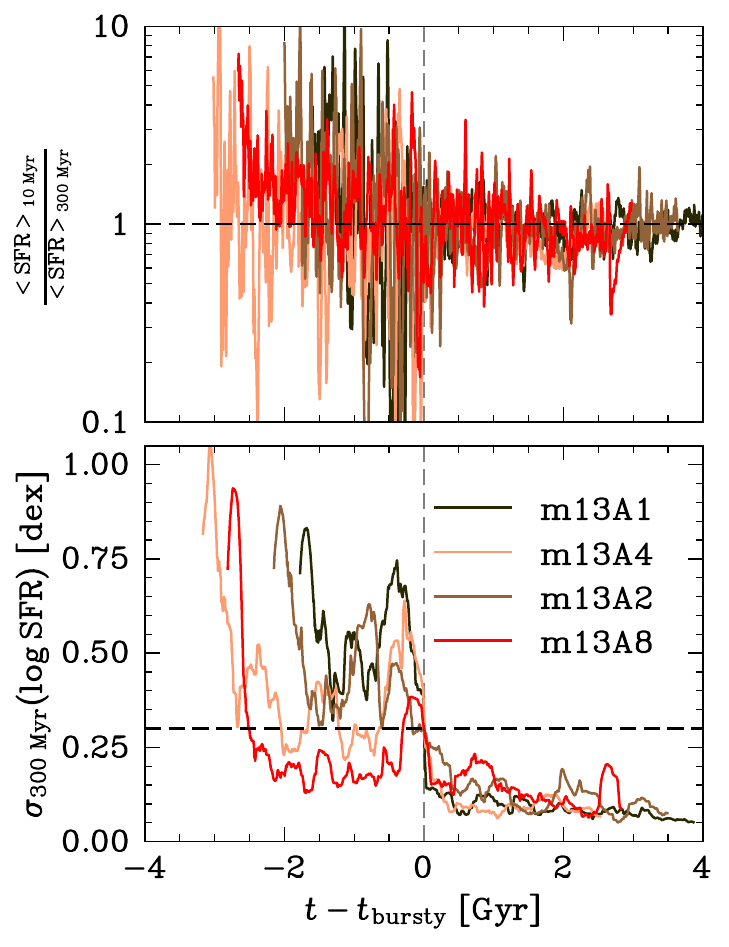}}
    \caption{Star formation histories for low-redshift (left) and high-redshift (right) halos in our sample. 
    The top panels show the instantaneous SFR normalized by the SFR averaged in a moving 300 Myr window.
    The bottom panels show the standard deviation of the SFR averaged in a moving 300 Myr window, $\sigma_{\mathrm{300\ Myr}}(\log\mathrm{SFR})$.
    The results are plotted as a function of cosmic time relative to $t_\mathrm{bursty}$, which is defined as the time when $\sigma_{\mathrm{300\ Myr}}(\log\mathrm{SFR})<0.3$.
    In both the low- and high-redshift regimes, massive halos undergo a rapid transition from bursty to steady star formation.
    }
    \label{fig:SFHscatter}
\end{figure*}

\subsection{Galaxy properties}\label{sec:galaxyprops}
Finally, we briefly revisit two galaxy-scale properties that have been the focus of previous FIRE studies, and compare trends across the high- and low-redshift halos in our sample.
In this section we explore fluctuations in the star formation rates of our galaxies, as well as the time evolution of rotational support in their ISM. 

\subsubsection{Bursty to steady star formation rates}\label{sec:burstytosteadySFR}
We measure the scatter in star formation rates as a function of cosmic time following the method described in Section 3.1 of \cite{gurvich_rapid_2023}.
Namely, we use the present-day masses and ages of star particles in the galaxy to construct an `archaeological' star formation history (SFH) (e.g., \citealt{floresvelazquezTimescalesProbedStar2021}).
We plot the ratio $\left<\mathrm{SFR} \right>_\mathrm{10\ Myr} / \left<\mathrm{SFR} \right>_\mathrm{300\ Myr}$, where the numerator and denominator represent the running mean of the SFH calculated in 10 Myr and 300 Myr windows, respectively.

Our results, which extend Figure 2 of \cite{gurvich_rapid_2023} to high-redshift massive galaxies, are shown in the upper panels of Figure \ref{fig:SFHscatter}.
In the the lower panels of the figure, we plot the standard deviation of the SFR averaged in a moving 300 Myr window, 
\begin{equation}
    \sigma_{\mathrm{300\ Myr}}(\log\mathrm{SFR}) = \sqrt{\left<(\log \mathrm{SFR})^2 \right>_\mathrm{300\ Myr} -\left<\log \mathrm{SFR} \right>^2_\mathrm{300\ Myr} }.
\end{equation}
We plot our results as a function of cosmic time relative to $t_\mathrm{bursty}$, which is defined as the \textit{last} time when $\sigma_{\mathrm{300\ Myr}}(\log\mathrm{SFR})<0.3$.
We use $t_\mathrm{bursty}$ as an indicator of when the star formation in a galaxies transitions from being `bursty' to more steady.

The three low-redshift MW-mass galaxies we analyze are shown in the left panels; these are the same three FIRE simulations analyzed by \cite{gurvich_rapid_2023}.
The MW-mass galaxies undergo a transition from bursty star formation at early times, characterized by large-amplitude fluctuations in their SFHs, to more steady star formation with significantly smaller fluctuations with time.
The more massive halos, shown in the right panels of Figure \ref{fig:SFHscatter}, behave similarly.
This is consistent with previous FIRE studies (e.g., \citealt{sternVirializationInnerCGM2021, byrneStellarFeedbackregulatedBlack2023, muratovGustyGaseousFlows2015, angles-alcazarCosmicBaryonCycle2017}) that found a transition in FIRE galaxies from bursty to more steady SFRs, roughly when the inner CGM virializes at a halo mass $M_\mathrm{vir} \sim 10^{12}\ \Msun$ (see Appendix \ref{app:tburstytICV}).

\subsubsection{Rotational support in the ISM}\label{sec:rotationsupportISM}
\begin{figure}
	\includegraphics[width=3.12in]{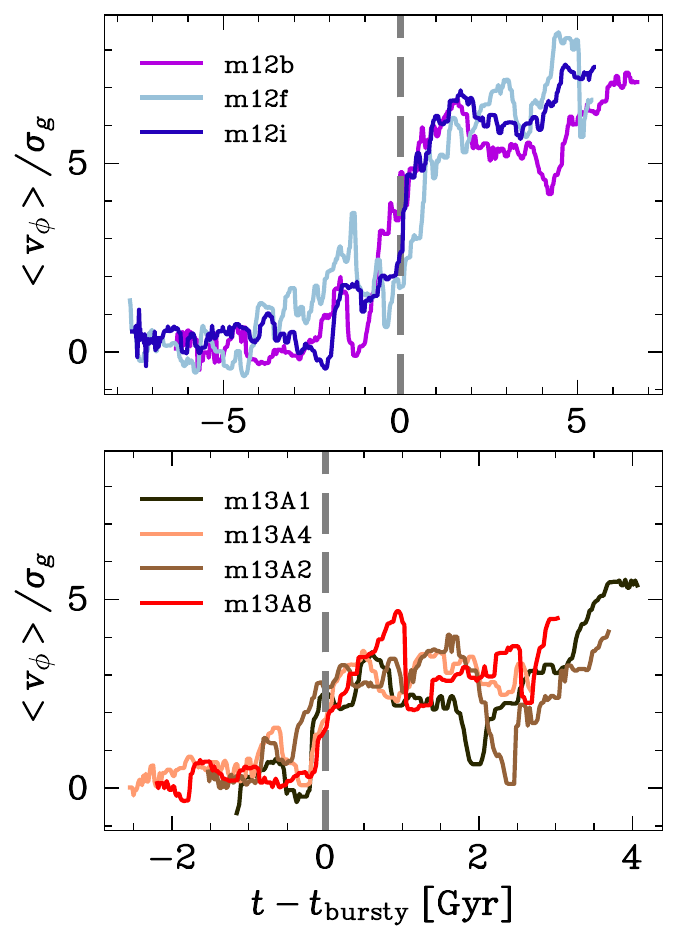}
    \caption{Time evolution of the ratio of the average azimuthal velocity $v_\phi$ of HI gas (calculated at $r<0.05 R_{\mathrm{vir}}$) to the dispersion of $v_\phi$ of HI gas.
    The upper panel shows the low-redshift MW-mass halos we analyze, while the lower panel shows the high-redshift more massive halos.
    The results are plotted as a function of cosmic time relative to $t_\mathrm{bursty}$, the time when the SFR becomes steady.
    The results are averaged over a moving 200 Myr window.
    At early times relative to $t_\mathrm{bursty}$, the ISM (traced by H I gas) is dispersion dominated and has negligible rotational support, with $\left< v_\phi \right>/\sigma_g \ll 1$.
    There is an increase in $\left< v_\phi \right>/\sigma_g$ at around $t_\mathrm{bursty}$, indicating the formation of a rotationally-supported disk.
    The low-redshift MW-mass halos reach higher ratios than the massive halos at higher redshifts, which are thicker and more turbulent.
    }
    \label{fig:vphi_sigma}
\end{figure}

Another galaxy property we analyze is the rotational support in the ISM, which is connected with the formation of galaxy disks.
We measure the ratio of the average azimuthal velocity to velocity dispersion of HI gas in the ISM ($r<0.05 R_{\mathrm{vir}}$), $\left< v_{\phi} \right>/\sigma_g$.
The HI-weighted average azimuthal velocity is $\left< v_{\phi} \right>$,
and we calculate the velocity dispersion of gas as $\sigma_g = \sqrt{\left< v_{\phi}^2 \right> - \left< v_{\phi} \right>^2}$.
$\left< v_{\phi} \right> /\sigma_g$ is a measure of the rotational support of the cool ISM H I gas, and large ratios imply a coherently rotating, cool disk.
In Figure \ref{fig:vphi_sigma}, we plot this ratio as a function of $t-t_\mathrm{bursty}$.
We median-average the results in a moving 200 Myr window to smooth over fluctuations at small time scales.

For both the m12 and m13 halos, at early times prior to $t_\mathrm{bursty}$, the ISM is dispersion dominated and has negligible rotational support ($\left< v_\phi \right>/\sigma_g \ll 1$).
At around the time of the bursty-to-steady star formation transition, $\left< v_\phi \right>/\sigma_g$ rapidly increases, indicating the formation of a rotationally-supported disk.
This is consistent with the result found by \cite{gurvich_rapid_2023} for MW-mass halos, and we show in Figure \ref{fig:vphi_sigma} that this trend holds for the more massive halos.
We note, however, that $\left< v_\phi \right>/\sigma_g$ can start increasing significantly before $t_\mathrm{bursty}$, corresponding to thick disk formation, which generally precedes the settling into thin disks more closely associated with $t_\mathrm{bursty}$ (see \citealt{gurvich_rapid_2023, yuBornThisWay2023, hopkinsWhatCausesFormation2023, byrneStellarFeedbackregulatedBlack2023}).

One important difference we find in the high redshift regime compared to lower redshifts is that gas disks are thicker and have higher amounts of dispersion.
As shown in Figure \ref{fig:vphi_sigma}, the massive high-redshift halos have $\left< v_\phi \right>/\sigma_g \lesssim 4$ for over 2 Gyr after $t_\mathrm{bursty}$, while the low-redshift halos reach $\left< v_\phi \right>/\sigma_g \gtrsim 5$ in this time.
We discuss this difference in Section \ref{sec:disc_lowz}.

We additionally checked the velocity dispersions of HI gas in the radial and polar velocities $v_r$ and $v_\theta$, and verified that while there are fluctuations in the dispersions as a function of time, $\sigma_g$ is also representative of the dispersions in $v_r$ and $v_\theta$. 
Furthermore, in comparison with the sound speed in $\sim10^4$~K gas traced by HI, the velocity dispersions are substantially supersonic.

\section{Results on the properties of cold and hot accretion on halo scales}\label{sec:ResultsI}
We start by analyzing gas inflows at the scale of the halo, using an Eulerian approach to measure instantaneous mass inflow rates and specific angular momentum.
In this section, we show the evolution of these properties over the $\sim$12.5 billion years from $z=5$ to $z=0$.

\subsection{Mass inflow rates}\label{sec:results_Mdot}

\begin{figure*}
	\includegraphics[width=6.5in]{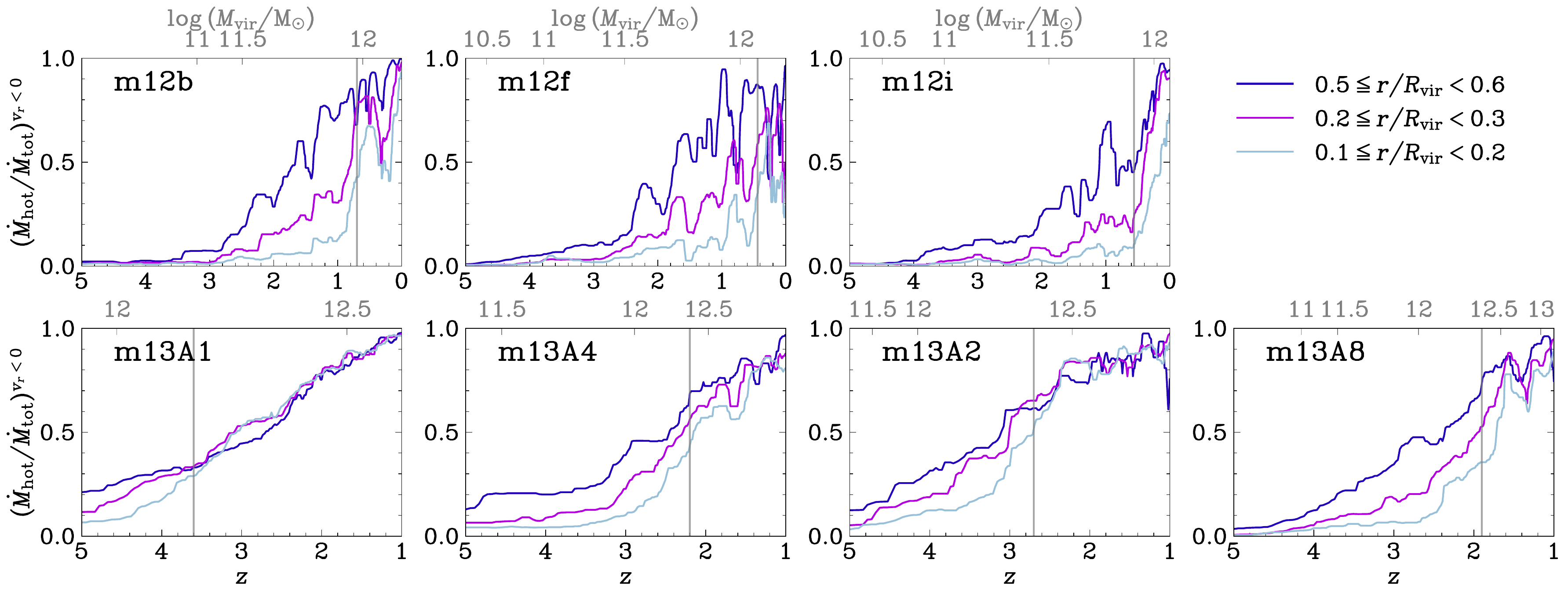}
    \caption{The hot gas inflow rate fractions in the CGM as a function of redshift.
    For the seven halos in our analysis set, ratios of the hot ($T > 10^5$ K) CGM mass inflow rate with respect to the total CGM inflow rate, $\left(\dot{M}_{\mathrm{hot}}/\dot{M}_{\mathrm{tot}}\right)^{v_r<0}$ are shown in three radial shells (indicated by different line colors).
    The halos are sorted by ascending order in $M_\mathrm{vir}(z=1)$.
    The results are averaged over a moving 400 Myr window.
    The halo virial mass in half-decade increments is shown at the top of each panel.
    The vertical lines mark $z_{\mathrm{ICV}}$, the redshift at which the inner CGM is expected to virialize.
    At ICV, the hot phase typically comprises $\sim50$\% of the inflow in the inner halo, and becomes increasingly dominant afterwards.
    On average, hot inflow fractions are larger at larger radius, consistent with higher gas densities in the inner halo, and outside-in CGM virialization. 
    }
    \label{fig:Mdotin_allz}
\end{figure*}

We first quantify instantaneous mass inflow rates of different gas phases.
At a given simulation snapshot, the mass flow rate of all inflowing gas within a radial shell $r_0 \leq r \leq r_1$ is given by
\begin{equation}
    \dot{M}_{\mathrm{tot}}^{v_r<0}=\frac{1}{\Delta r} \sum_{v_{r,i}<0} v_{r,i} m_i
\end{equation}
where $\Delta r = r_1 - r_0$ is the width of the shell,
and $v_{r,i}$ and $m_i$ are the radial velocity and mass of a gas particle, respectively.
To quantify gas inflows while excluding outflowing gas, we sum over all particles $i$ within the shell for which $v_{r,i}<0$.
In gas with a large velocity dispersion, our radial velocity cut could bias our measured inflow rate by selecting half of the turbulent component rather than gas that is truly inflowing. 
To test for this effect, we repeated our analysis using the net mass flow without imposing the $v_r<0$ cut. 
The results below remain similar in magnitude (although the net hot gas fraction falls to $\sim 0$ at certain times due likely to strong outflows) indicating that our results in this section are not significantly biased by this effect.

In addition to finding the total inflow rate $\dot{M}_{\mathrm{tot}}^{v_r<0}$, we distinguish gas accretion in the hot and cold modes by restricting our measurement to hot ($T_i > 10^5$ K) and cool ($T_i < 10^5$ K) particles, where $T_i$ is the gas temperature.
A large number of previous simulation studies of gas accretion in halos have used the full thermodynamic histories of particles to identify hot and cold accretion, typically by applying a threshold to the maximum past temperature $T_\mathrm{max}$ to select particles that were heated by an accretion shock (e.g., \citealt{keresHowGalaxiesGet2005, keresGalaxiesSimulatedLCDM2009, vandevoortRatesModesGas2011, nelsonMovingMeshCosmology2013}).
While much of this work was based on simulations without galactic outflows, modern simulations implement strong outflows in order to match various observations (see reviews by, e.g., \citealt{somervillePhysicalModelsGalaxy2015, naabTheoreticalChallengesGalaxy2017a}). In such simulations, shocks in CGM gas are frequently caused not only by accretion, but also by feedback-driven outflows.
In this study, our aim is not to determine whether gas was shock heated in the past, but whether shock-heated gas remains hot or cools rapidly.
For many purposes, such as assessing whether the hot gas is long lived or dynamically important, this is the more directly relevant question.
We therefore use the \textit{instantaneous} gas temperature rather than $T_\mathrm{max}$ to define hot and cold accretion, and adopt $10^5$~K as a proxy for separating rapidly cooling gas from gas in the hot phase.
As shown by, e.g., \cite{sultanCoolingFlowsReference2025}, this threshold approximately separates cool gas from hot, virialized gas (see also \citealt{faucher-giguereBaryonicAssemblyDark2011}).

In Figure \ref{fig:Mdotin_allz}, we show hot-gas inflow rate fractions, defined as the ratio of the mass inflow rate of hot gas, $\dot{M}_{\mathrm{hot}}^{v_r<0}$, to the total inflow rate. 
Results for $\left(\dot{M}_{\mathrm{hot}}/\dot{M}_{\mathrm{tot}}\right)^{v_r<0}$ are shown for the halos in our sample from $z=5$ to the final snapshot of the simulation, which is $z=0$ for the low-redshift MW-mass halos (first row of the figure), and $z=1$ for the high-redshift more massive halos (second row).
To smooth fluctuations at small time scales, we median-average the hot inflow rate fractions in a moving 400 Myr window.

We show results in three radial shells with inner radii of 0.1, 0.2, and $0.5R_\mathrm{vir}$, each with a shell thickness of $0.1R_\mathrm{vir}$.
At redshift $z \sim 5$, cool gas dominates the inflow in all three shells.
The inflows are increasingly made up of hot gas as redshift decreases, and by the final redshifts shown, the hot mode generally dominates the inflow.
Note that two of the halos, m12i and m12f, develop high hot-mode inflow rate fractions by $z\sim0$, but contain more cold inflowing gas at inner radii:
in the inner CGM of these two halos, $\left(\dot{M}_{\mathrm{hot}}/\dot{M}_{\mathrm{tot}}\right)^{v_r<0} \lesssim 0.7$ at $z=0$.
As discussed in Section 5.3 of \cite{sultanCoolingFlowsReference2025}, the large fractions of cool inflowing gas in halos at this mass scale and redshift may have cooled out of the hot phase due to nonlinear thermal instabilities (see also, \citealt{esmerianThermalInstabilityCGM2021}).

The evolution from cold to hot mode accretion across our halos is a signature of the virialization of their gas.
The vertical lines in Figure \ref{fig:Mdotin_allz} indicate $z_{\mathrm{ICV}}$.
Focusing on the inner CGM shell at $0.1R_\mathrm{vir}$, the hot-inflow rate fractions are nearly zero well before $z_{\mathrm{ICV}}$.
As expected, most of the halos have an increase in the hot-inflow rate fraction in the inner shell at $z_{\mathrm{ICV}}$, and well after this redshift, are dominated by hot mode accretion.

Leading up to $z_{\mathrm{ICV}}$, in all of the halos we analyzed, at fixed redshift $\left(\dot{M}_{\mathrm{hot}}/\dot{M}_{\mathrm{tot}}\right)^{v_r<0}$ increases with increasing radius.
This demonstrates that the halo undergoes the change from hot to cold mode accretion in an outside-in fashion, which is consistent with the ICV scenario found by \cite{sternVirializationInnerCGM2021}.
As we show in Appendix \ref{app:Mdotprofiles}, the radial gradient in $\left(\dot{M}_{\mathrm{hot}}/\dot{M}_{\mathrm{tot}}\right)^{v_r<0}$ is primarily driven by the hot gas inflow rate $\dot{M}_{\mathrm{hot}}^{v_r<0}$, which generally increases with increasing radius (see Figure \ref{fig:Mdotprofiles} for radial profiles of mass inflow rates).

Figure \ref{fig:Mdotin_allz} demonstrates the gradual evolution that halos undergo in their accretion as their inner CGM virializes at $M_\mathrm{vir}\sim10^{12}\ \Msun$.
Their inflows are almost entirely cool $\gtrsim$ Gyr prior to ICV, while hot gas dominates the inflow at times $\gtrsim$ Gyr after virialization of the inner halo.

\subsection{Angular momentum}\label{sec:results_ang_mom}

\begin{figure*}
	\includegraphics[width=6.5in]{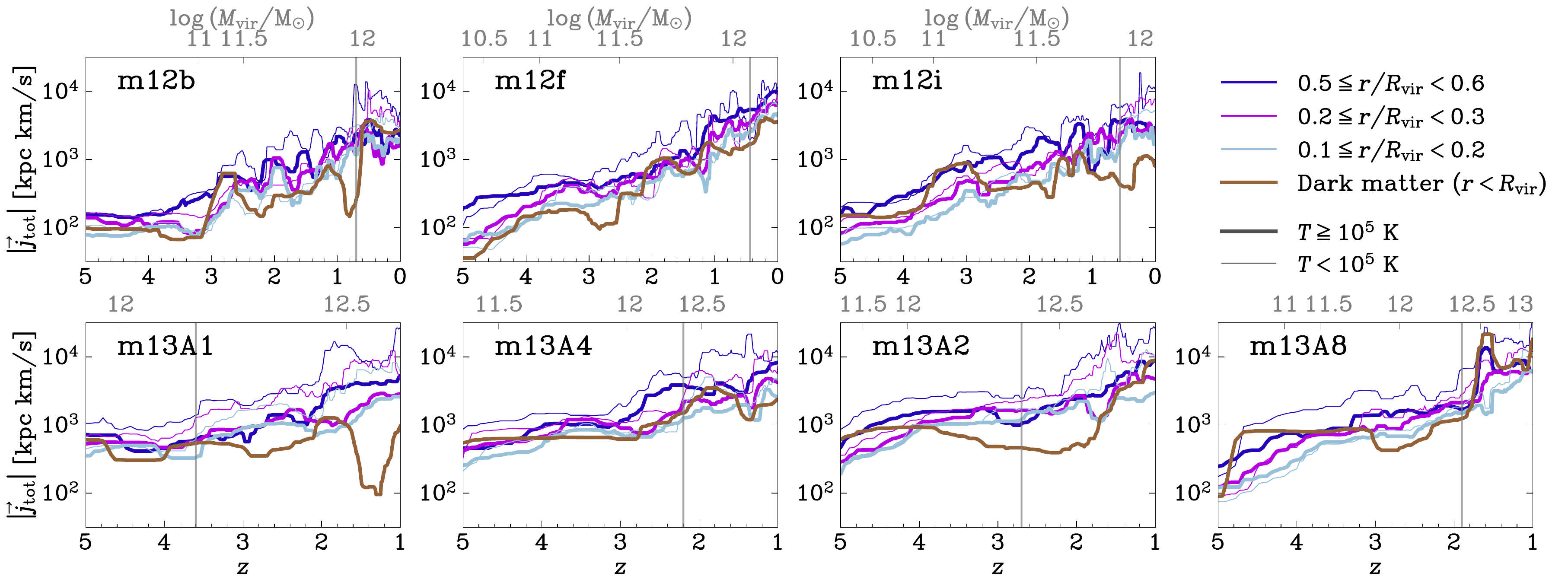}
    \caption{The magnitude of the total specific angular momentum of hot and cold inflowing gas in the CGM as a function of redshift.
    For the seven halos in our analysis set, $|\vec{j}_\mathrm{tot}|=|\vec{J}_\mathrm{tot}/M_\mathrm{tot}|^{v_r<0}$ is shown for hot ($T \ge 10^5$ K; thick lines) and cold ($T < 10^{5}$ K; thin lines) CGM gas with $v_r<0$.
    The thick brown line shows the specific angular momentum of all dark matter particles in the halo. 
    The results are averaged over a moving 400 Myr window, and shown in three radial bins (indicated by different line colors).
    The vertical lines mark $z_{\mathrm{ICV}}$, the redshift at which the inner CGM is expected to virialize.
    The halo virial mass in half-decade increments is shown at the top of each panel.
    Cold inflows generally carry the highest amounts of specific angular momentum, followed by hot inflows and dark matter, consistent with previous simulation studies.
    }
    \label{fig:jtot_allz}
\end{figure*}

\begin{figure*}
	\includegraphics[width=6.5in]{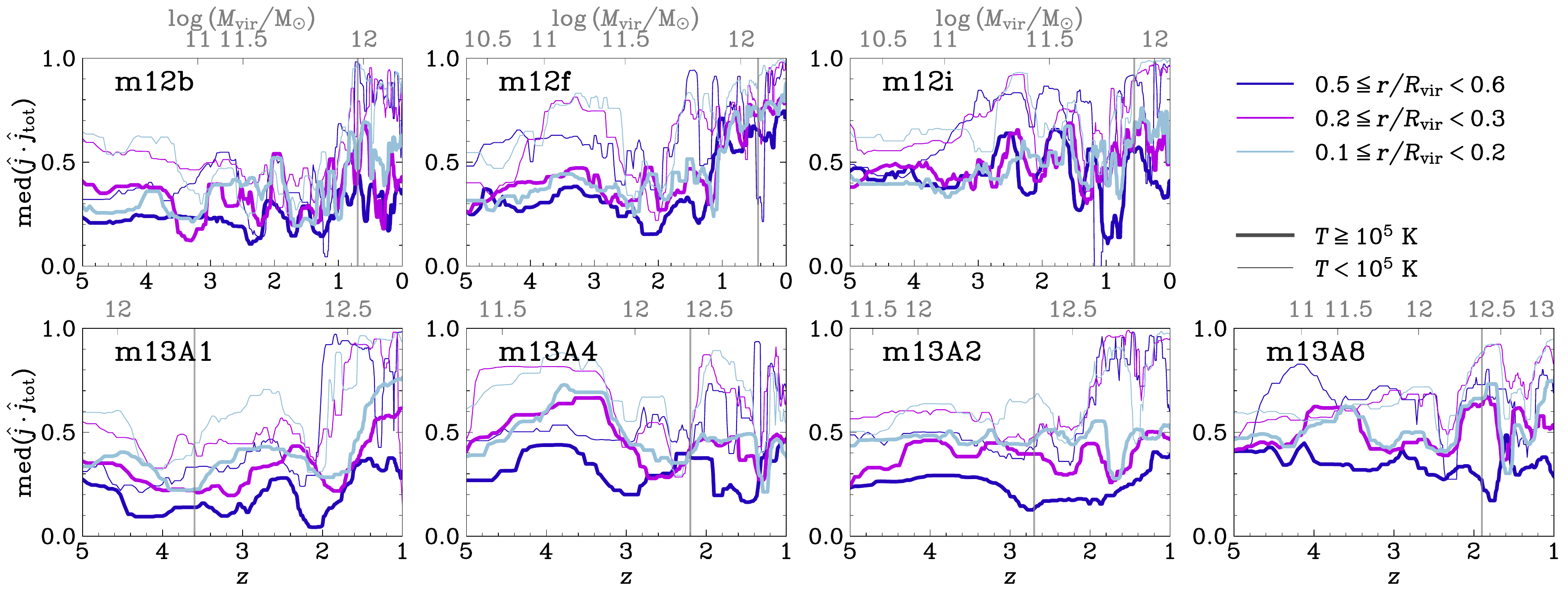}
    \caption{The coherence of the specific angular momentum of hot and cold inflowing gas in the CGM as a function of redshift.
    Similar to Figure \ref{fig:jtot_allz}, this figure shows the median value of $\hat{j} \cdot \hat{j}_\mathrm{tot}$, calculated separately for hot ($T \ge 10^5$ K; thick lines) and cold ($T < 10^{5}$ K; thin lines) CGM gas with $v_r<0$.
    The results are averaged over a moving 400 Myr window, and shown in three radial bins (indicated by different line colors).
    The cold inflows are more coherent in their angular momentum than the hot inflows, at the radii probed here.
    }
    \label{fig:jcohere_allz}
\end{figure*}

In addition to measuring the mass flow rates of inflowing gas, we explore how angular momentum is carried by the hot and cold inflows.
The specific angular momentum of a gas particle $i$ is $\vec{j}_i=\vec{r}_i \times \vec{v}_i$.
We define the total specific angular momentum of the inflowing gas particles in a radial shell as
\begin{equation}\label{eq:jtot}
    \vec{j}_\mathrm{tot} = \frac{\sum_{v_{r,i}<0} m_i \vec{r}_i \times \vec{v}_i}{\sum_{v_{r,i}<0} m_i},
\end{equation}
where the numerator on the right hand side is the total angular momentum of the shell, $\vec{J}_\mathrm{tot}^{v_r<0}$, and the denominator is the total mass, $M_\mathrm{tot}^{v_r<0}$.
Here, $m_i$ is the mass of the particle, and $\vec{r}_i$ and $\vec{v}_i$ are the position and velocity vectors of the particle relative to the center of the halo (see Section \ref{sec:methods_halocentering}).

In Figure \ref{fig:jtot_allz}, we show the angular momentum belonging to different gas phases by calculating Equation \ref{eq:jtot} for hot ($T \ge 10^5$ K) and cold ($T < 10^{5}$ K) particles, separately.
The results are shown for the same three radial shells as in Figure \ref{fig:Mdotin_allz}, encompassing both the inner and outer parts of the halo, and we median-average the results in a moving 400 Myr window to smooth over fluctuations. 
The hot and cold inflows are indicated by the thick and thin lines, respectively.
The total specific angular momentum (sAM) throughout the halo generally increases (or remains constant) with cosmic time, which is apparent for both the hot and cold inflow.

Comparing our results between the gas phases, we generally measure greater amounts of sAM carried by cold inflows than by hot inflows.
We also measure the specific angular momentum of all dark matter within the virial radius, indicated by the thick brown line in Figure \ref{fig:jtot_allz}.
The gas generally contains more sAM than the dark matter. 
This is consistent with previous studies across a wide range of simulations, as we discuss in Section \ref{sec:disc_angmom}.

We additionally measure the average coherence of the specific angular momentum of the hot and cold phases.
An inflow with high specific angular momentum does not guarantee high net AM delivery to the galaxy if the flow is highly incoherent (and thus leading to large cancellations in its AM vector as it accretes), so the inflow must be both high in sAM and coherent to contribute sAM to the galaxy.
We quantify this in a given radial shell, and for hot and cold inflowing particles separately, as the median value of 
\begin{equation}
    \hat{j}_i \cdot \hat{j}_\mathrm{tot,\ phase} = \frac{\vec{j}_i \cdot \vec{j}_\mathrm{tot,\ phase}}{|j_i| |j_\mathrm{tot,\ phase}|} = \cos \theta_i.
\end{equation}
Here, $\vec{j}_\mathrm{tot,\ phase}$ is the total sAM of the hot or cold inflow as defined by Equation \ref{eq:jtot}.
Defined in this way, the average coherence is equivalent to the median cosine of the angle $\theta_i$ between the sAM vector of a particle and the total sAM of all particles, and ranges from 1 (exactly aligned) to -1 (exactly anti-aligned).
A coherence of 0 ($\vec{j}_i \cdot \vec{j}_\mathrm{tot,\ phase}=0$) indicates a lack of alignment.

The median coherence of hot and cold inflowing gas are displayed in Figure \ref{fig:jcohere_allz}, measured in the same radial shells as the earlier results presented in this section.
The cold inflow is generally more coherent than the hot inflow in a given radial shell.
Although the coherence of the inflows often fluctuates and does not monotonically increase with time, generally the inflows are more coherent at late times than at early times.
In most cases, the cold inflow in the inner radial shell reaches a high alignment of $\gtrsim0.8$ by the last snapshot.
The hot inflow exhibits low amounts of coherence, and generally $\mathrm{med}(\hat{j}_i \cdot \hat{j}_\mathrm{tot}) \lesssim 0.5$.

Figures \ref{fig:jtot_allz} and \ref{fig:jcohere_allz} show that cold inflows generally carry more specific angular momentum than hot inflows, in agreement with many previous simulation studies (e.g., \citealt{danovichFourPhasesAngularmomentum2015, stewartHighAngularMomentum2017, defelippisAngularMomentumCircumgalactic2020, wangLargescaleEnvironmentCGM2021}).
Cold inflows are also more coherent in their sAM than hot inflows, as found by \cite{danovichFourPhasesAngularmomentum2015}.
At face value, this may seem to contradict, e.g., \cite{hafenHotmodeAccretionPhysics2022}, which found that in post-ICV halos, the hot mode becomes coherently rotating prior to accretion and is the dominant contributor of AM to the galaxy.
However, as we show in the following section, the timescale for circularization of a typical hot gas particle is very short ($\lesssim100$ Myr) and occurs within the innermost halo ($r \lesssim0.1 R_\mathrm{vir})$), so this important effect is not apparent in the CGM diagnostics of this section.

\begin{figure*}
	\includegraphics[width=6.5in]{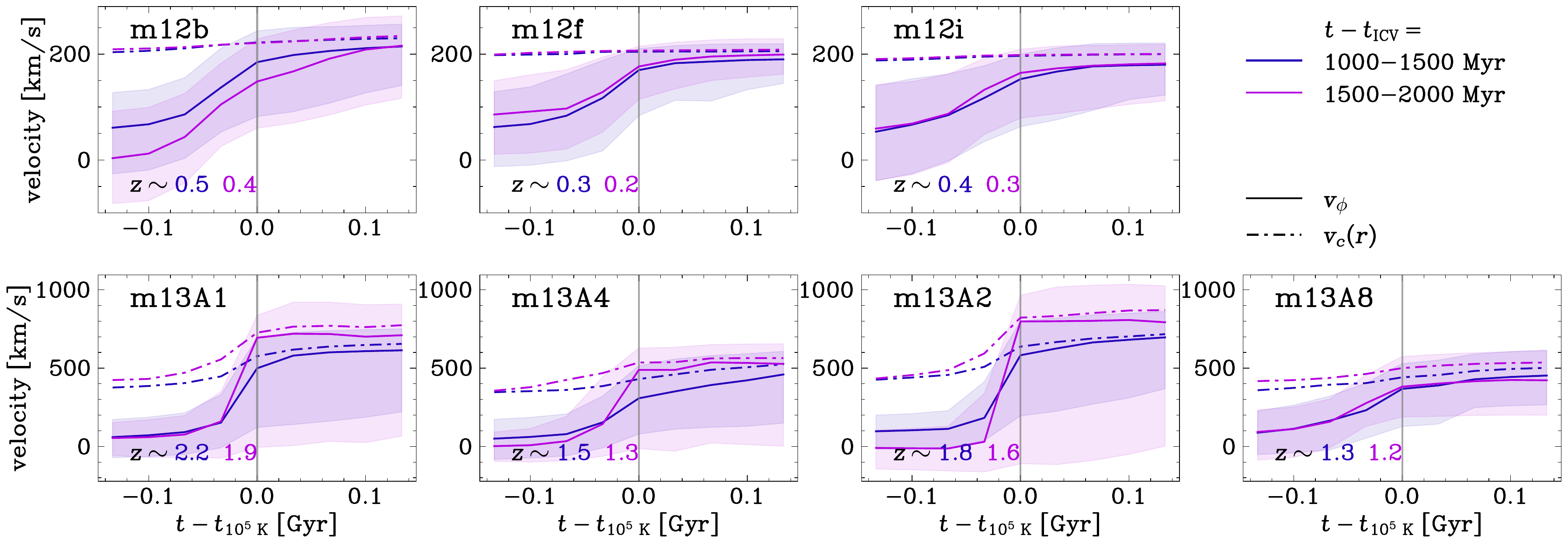}
    \caption{Tracks of azimuthal velocity of accreted gas particles, shown for particles that are initially in the hot phase and are accreted during two time bins subsequent to ICV.
    Tracks are shown with respect to $t_\mathrm{10^5\ \mathrm{K}}$, which is the time of the first snapshot in the time bin for which $T<10^5$ K.
    The solid lines indicate median tracks, while the shaded bands represent the 16- to 84- percentiles.
    The dot-dashed line shows the median circular velocity track.
    The azimuthal velocity approaches $v_c$ as it cools, indicating the inflows have circularized.
    Accreting particles that do not cool below $10^5$ K during the time window are excluded.
    The median redshift of the 500 Myr tracking windows are listed in each panel.
    In post-ICV halos, hot inflowing gas circularizes simultaneously with cooling.
    }
    \label{fig:vphi_ICV}
\end{figure*}

\begin{figure*}
	\includegraphics[width=6.5in]{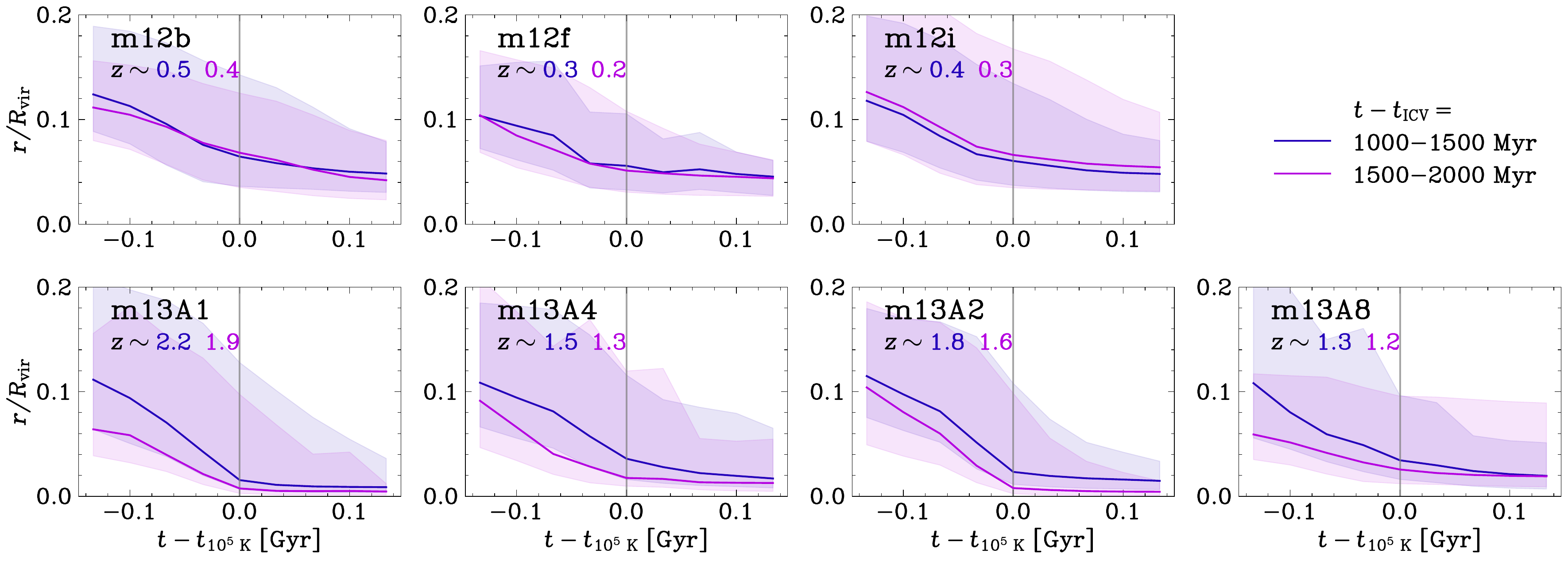}
    \caption{Tracks of the radius of accreted gas particles, shown for particles that are initially in the hot phase.
    Tracks are shown for hot gas accreted during two time bins subsequent to ICV, as in Figure \ref{fig:vphi_ICV}, at which accretion from the hot phase is dominant.
    Tracks are shown with respect to $t_\mathrm{10^5\ \mathrm{K}}$, which is the time of the first snapshot in the time bin for which $T<10^5$ K.
    The solid lines indicate median tracks, while the shaded bands represent the 16- to 84- percentiles.
    Accreting particles that do not cool below $10^5$ K during the time window are excluded.
    Hot gas cools at $r< 0.1 R_\mathrm{vir}$, and stalls at galactic radii of $r\sim 0.01-0.05 R_\mathrm{vir}$.
    }
    \label{fig:rscaled_ICV}
\end{figure*}

\section{Results on the physics of gas accretion onto the central galaxy}\label{sec:ResultsII}
We now analyze the properties of the inflows as they approach the galaxy.
In contrast to the instantaneous Eulerian mass inflow rates and angular momentum results we presented above in our halo-scale analysis, in this section we utilize particle tracking to gain more insights on the properties of the gas as it inflows through the inner halo and accretes onto the galaxy.

\subsection{Inflow circularization relative to cooling times}\label{sec:results_hotparticletracking}
As described below in Section \ref{sec:methods_particletracking}, we explicitly track the Lagrangian particles in the meshless simulations that inflow from the CGM ($0.1 R_{\mathrm{vir}} <r< R_{\mathrm{vir}}$) to galaxy scales ($r<0.05R_{\mathrm{vir}}$).
In this analysis, we focus on the hot accretion mode and analyze halos that have completed virialization of their inner CGM.

\subsubsection{Tracking accreting hot-phase gas particles}\label{sec:methods_particletracking}
We track particles that accrete onto the central galaxy using the method described below and similar (but not identical) to the procedure carried out by \cite{hafenHotmodeAccretionPhysics2022}.
We track particles accreted between redshifts $z_0$ and $z_1$, where $z_0 > z_1$, by finding all particles that meet the following criteria: (i) at $z_0$, the particle is a gas particle in the CGM ($0.1 R_{\mathrm{vir}} <r<R_{\mathrm{vir}}$) that belongs to the hot phase, (ii) at $z_1$, the particle is inside the galaxy ($r<0.05 R_{\mathrm{vir}}$), and is either a gas or star particle.
We identify particles belonging to the hot phase at $z_0$ using the virial-branch selection algorithm described in Appendix A of \cite{sultanCoolingFlowsReference2025}.
Briefly, we make a radius-dependent selection in temperature and density space of the dominant phase of gas with $T>10^5$ K.

After finding all particles that meet the above criteria, we track the properties of the particles (e.g., temperature, density, and 3D velocity) at every simulation snapshot between $z_0$ and $z_1$.
We rotate particle coordinates and velocities at each snapshot such that the z-axis is the axis of rotation of the central galaxy at $z_1$.
We define the rotation axis as described in Section \ref{sec:methods_halocentering}.

We are interested in how the properties of particles evolve with respect to their time of cooling, $t_\mathrm{10^5\ \mathrm{K}}$.
We define $t_\mathrm{10^5\ \mathrm{K}}$ as the time associated with the first snapshot (in the redshift range $z_0 \le z \le z_1$) for which $T<10^5$ K.

\subsubsection{Evolution of rotational velocities}
In Figure \ref{fig:vphi_ICV}, we show tracks of the azimuthal velocities of accreting gas particles that were initially in the hot phase.
Results are shown for particles that enter within $r=0.05R_{\mathrm{vir}}$ during two time bins.
The time bins begin 1000 and 1500 Myr after ICV and are each 500 Myr long.
As shown in Section \ref{sec:results_Mdot}, inflows in the halos we analyze become dominated by hot gas as the inner CGM virializes.  
Thus, the time bins probe the post-virialization hot mode-dominated halos of steady galaxies.

The solid lines indicate median tracks of the particles, while the shaded bands represent the 16- to 84- percentiles.
We show our tracks relative to $t_\mathrm{10^5\ \mathrm{K}}$, which is the time of the first snapshot in the time bin for which a particle cools below $10^5$ K.
Accreting hot-phase particles that do not cool below $10^5$ K during the time bin are excluded from our results.
The dot-dashed line shows the median circular velocity, which for a gas particle at a distance $r$ from the center of the halo is given by $v_c \equiv \sqrt{\frac{G M(<r)}{r}}$.

At early times relative to the particles' cooling time $t_\mathrm{10^5\ \mathrm{K}}$, most particles have low rotational velocities, with $v_\phi \ll v_c$ in the median.
There is a rapid increase in rotational velocity which begins $\lesssim 50$ Myr prior to cooling.
By $t_\mathrm{10^5\ \mathrm{K}}$ or shortly afterwards, the median rotational velocity stabilizes at $v_\phi \sim v_c$.
This trend, which is consistent across the seven halos in our sample in both time bins we explore, suggests that in post-ICV halos, hot inflowing gas circularizes as it cools.
This is consistent with the previous analysis of low-redshift, Milky Way-mass halos in FIRE by \cite{hafenHotmodeAccretionPhysics2022} and with the more idealized rotating cooling flow models from \cite{sternAccretionDiscGalaxies2024}. 
These previous studies showed that cooling is induced by circularization, as a forming disk increases gas densities and angular momentum support halts the inflow. 
By the it cools to $10^{5}$ K, inflowing has already become almost entirely rotationally supported.
The significant spread in rotational velocities after cooling in some cases (m13A1 and m13A2) may reflect the presence of turbulence in the ISM. 
Additionally, differences in the ordering of the curves corresponding to the two time bins between the panels in Figure \ref{fig:vphi_ICV} likely reflect stochastic variability across the simulations.

Turbulence may also be the primary reason why the median $v_\phi$ tracks for cool gas (i.e., when $t-t_\mathrm{10^5\ \mathrm{K}}>0$) are systematically slightly below the circular velocity of the gravitational potential.
Observational studies of disk galaxies such as \cite{kassinStellarMassTullyFisher2007} found that scatter in the Tully-Fisher relation is significantly reduced when velocity dispersion is added in quadrature with rotational velocity. 
\cite{wellonsMeasuringDynamicalMasses2020} analyzed most of the same FIRE m12 and m13 simulated galaxies we analyze in this work and similarly found that turbulent pressure gradients significantly contribute to radial force balance in 
galactic disks. 
Thus, velocity dispersion must be taken into account to correctly infer dynamical masses from gas rotational velocities (sometimes called the ``asymmetric drift'' correction).

\subsubsection{Cooling radii}
Next we show tracks of radial distance of the accreting hot-mode particles analyzed in the previous section.
Figure \ref{fig:rscaled_ICV} shows the median $r/R_\mathrm{vir}$ tracks of the particles; the shaded bands indicate the 16- to 84- percentiles.
As in Figure \ref{fig:vphi_ICV}, tracks are plotted with respect to $t_\mathrm{10^5\ \mathrm{K}}$ for particles that accrete onto the galactic radius in two post-ICV time bins, and particles that do not cool below $10^5$ K during the time bin are excluded.

Figure \ref{fig:rscaled_ICV} reveals that $\sim$150 Myr prior to cooling, most particles reside in the inner CGM.
The tracks of radial distance decrease until $t \sim t_\mathrm{10^5\ \mathrm{K}}$, at which time $r\sim 0.05 R_\mathrm{vir}$ for the low-redshift MW-mass halos, and $r\sim 0.02 R_\mathrm{vir}$ for the high-redshift massive halos.
At around this time, the median tracks stabilize.
As shown in Figure \ref{fig:vphi_ICV}, by $t \sim t_\mathrm{10^5\ \mathrm{K}}$ the inflow becomes rotationally supported.
Thus, the inflow stalls at the circularization radius, and the tracks of radial distance stabilize.

Figures \ref{fig:vphi_ICV} and \ref{fig:rscaled_ICV} together demonstrate that in post-ICV halos, hot inflowing gas rapidly circularizes at $r \lesssim 0.1R_\mathrm{vir}$ over a short timescale (of $\lesssim100$ Myr) before it cools.
This explains why a coherent hot phase is not evident in Figure \ref{fig:jcohere_allz}, which includes larger radii.

\begin{figure*}
	\includegraphics[width=6.5in]{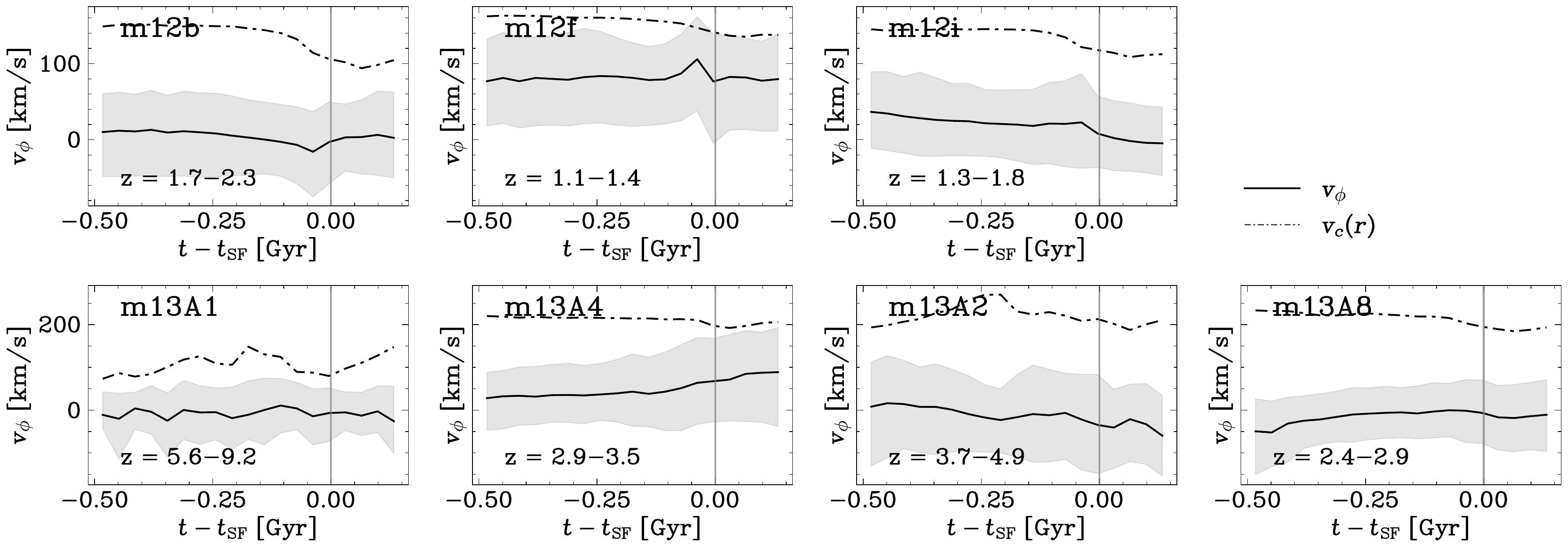}
    \caption{Tracks of azimuthal velocity of accreted gas particles, shown for particles that are initially in the cold phase and are accreted during an early time bin prior to ICV.
    Tracks are shown with respect to $t_\mathrm{SF}$, defined as the time of the first snapshot in which a gas particle forms a star.
    We select cold gas particles at $0.3 R_{\mathrm{vir}} <r<0.5R_{\mathrm{vir}}$ at 4.5 Gyr (1.25 Gyr) before ICV for m12 (m13) halos, track them for 1 Gyr (500 Myr), and require that they are stars within the galaxy by the end of this interval.
    The redshift range over which particles are tracked is indicated in each panel.
    The tracking windows are well before ICV, when all of the galaxies are highly bursty.
    The solid lines indicate median tracks, while the shaded bands represent the 16- to 84- percentiles.
    The dot-dashed line shows the median circular velocity track.
    In bursty galaxies, cold inflowing gas that dominates accretion in the inner CGM does not circularize prior to star formation.
    }
    \label{fig:vphi_tstarform_earlytimes}
\end{figure*}

\begin{figure*}
	\includegraphics[width=6.5in]{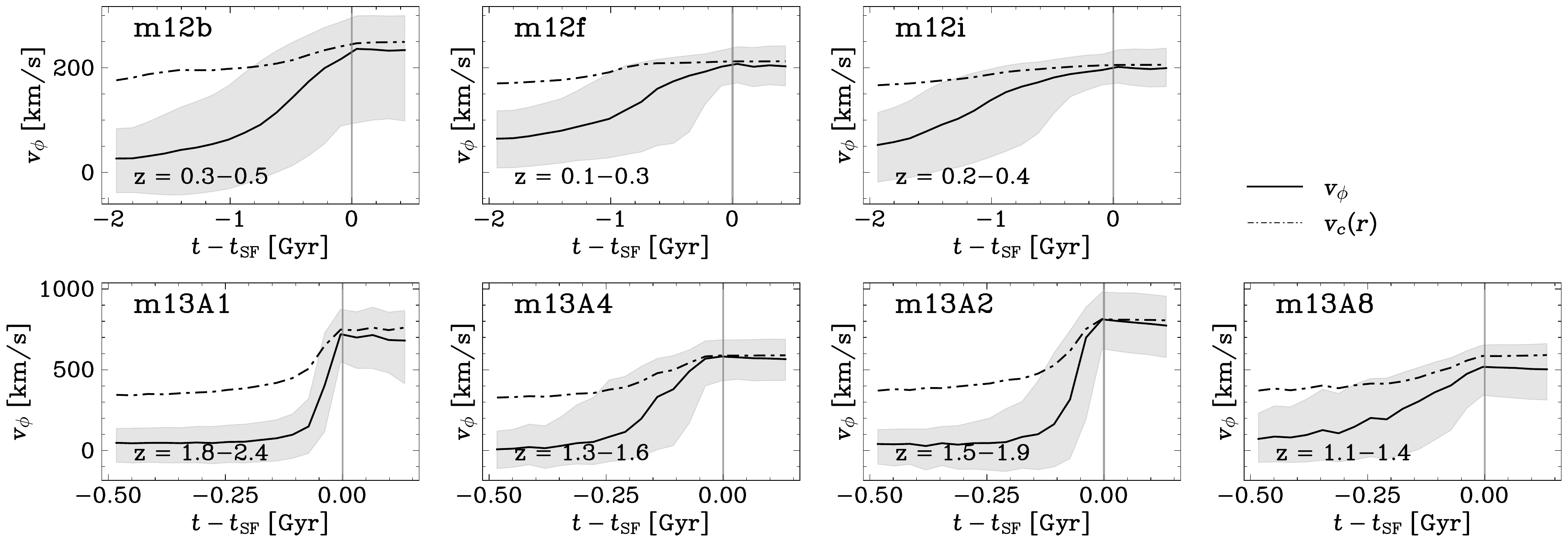}
    \caption{Tracks of azimuthal velocity of accreted gas particles, shown for particles that are initially in the hot phase and are accreted during a late time bin following ICV.
    Tracks are shown with respect to $t_\mathrm{SF}$, defined as the time of the first snapshot in which a gas particle forms a star.
    We select hot gas particles at $0.3 R_{\mathrm{vir}} <r<0.5R_{\mathrm{vir}}$ at 1 Gyr after ICV, track them for 2 Gyr (1 Gyr) for the m12 (m13) halos, and require that they are stars within the galaxy by the end of this interval.
    The redshift range over which particles are tracked is indicated in each panel.
    The tracking windows are after ICV, when the galaxies are steady.
    The solid lines indicate median tracks, while the shaded bands represent the 16- to 84- percentiles.
    The dot-dashed line shows the median circular velocity track.
    In steady galaxies with a virialized halo, hot inflowing gas that dominates accretion in the CGM generally circularizes rapidly before star formation.
    }
    \label{fig:vphi_tstarform_latetimes}
\end{figure*}

\subsection{Inflow circularization relative to star formation times}\label{sec:results_parttracking_tform}
Next, we track inflowing gas particles that eventually form stars and study whether, and when, the inflows circularize relative to the time of star formation, $t_\mathrm{SF}$.
Our method is similar to the particle tracking procedure we use in Section \ref{sec:results_hotparticletracking}, but here we measure azimuthal velocity tracks with respect to $t_\mathrm{SF}$, defined as the time of the first snapshot in which a gas particle becomes a star particle.
We consider halos in both the cold- and hot-mode accretion regimes.

We first analyze halos at times significantly before ICV, when inflows are dominated by cold gas and the central galaxies have bursty star formation.
Over a time window spanning redshifts $z_0 > z_1$, we track particles that meet the following criteria: (i) at $z_0$, the particle is a gas particle in the middle halo ($0.3 R_{\mathrm{vir}} <r<0.5R_{\mathrm{vir}}$) that belongs to the cold phase ($T<10^5$ K), and (ii) at $z_1$, the particle is inside the galaxy ($r<0.1 R_{\mathrm{vir}}$) and is a star particle.
We choose window lengths of 1 Gyr for the m12 halos and 500 Myr for the m13 halos that, as discussed below, allow roughly sufficient time for particles in our selection to form stars.
Similarly, we initially consider particles at $0.3 R_{\mathrm{vir}} <r<0.5R_{\mathrm{vir}}$ in order to allow enough time for particle accretion through the inner halo and subsequent star formation during the tracking window.

We carry out a parallel analysis at late times relative to ICV, when inflows are dominated by hot gas and the central galaxies are steady.
Over a 2 Gyr (1 Gyr) time window starting 1 Gyr after ICV for the m12 (m13) halos, we track particles that are initially gas in the middle halo ($0.3 R_{\mathrm{vir}} <r<0.5R_{\mathrm{vir}}$) and belong to the hot phase (see Section \ref{sec:methods_particletracking} for our hot phase selection), and that by the end of the tracking window are star particles inside the galaxy ($r<0.1 R_{\mathrm{vir}}$).

In Figure \ref{fig:vphi_tstarform_earlytimes}, we show the median tracks of azimuthal velocity for cold particles accreted in bursty galaxies (4.5-3.5 Gyr before ICV for the m12 halos, and 1.25-0.75 Gyr before ICV for the m13 halos).
Generally, particles that are initially cold have low rotational velocities that remain small before and after star formation, with $v_\phi < v_c$ in the median.
The cold inflowing gas that dominates accretion in halos at times well before ICV shows no clear signature of circularization prior to star formation.
We also find that $\sim10-45\%$ of the initially cold particles are temporarily heated above $10^5$ K for some period prior to star formation.
Although circular velocity generally increases as radius decreases from inner CGM to ISM scales, the median $v_c$ tracks shown are not strictly monotonic in time since the particles in each $t-t_\mathrm{SF}$ bin span a range of radii.

Figure \ref{fig:vphi_tstarform_latetimes} shows the analogous results for hot-mode particles accreted in steady galaxies.
The behavior in this regime is dramatically different.
In most of the halos, hot gas does not rotate initially ($v_\phi \ll v_c$) and becomes rotationally supported before star formation.

A potential bias in our particle selection in Figures \ref{fig:vphi_tstarform_earlytimes} and \ref{fig:vphi_tstarform_latetimes} is that gas forming stars on timescales longer than the tracking window is excluded.
To counteract this, we use tracking windows of 1 Gyr (500 Myr) for bursty galaxies and 2 Gyr (1 Gyr) for steady galaxies for the m12 (m13) halos, chosen to allow enough time for most accreted gas to form stars.
These choices are motivated by our analysis in Section \ref{sec:results_residencetimes} of the time accreted particles resides in bursty and steady galaxies before forming stars.
The timescales on which hot inflows circularize before star formation that we find in Figure \ref{fig:vphi_tstarform_latetimes} ($1-2$ Gyr for the m12 halos and $100-500$ Myr for the m13 halos) are roughly consistent with the residence time measurements we present in Section \ref{sec:results_residencetimes}.

Figures \ref{fig:vphi_tstarform_earlytimes} and \ref{fig:vphi_tstarform_latetimes} show that the circularization of inflowing gas strongly depends on the state of the CGM.
Cold-mode inflows in pre-ICV halos form stars without developing significant rotational support, while hot-mode inflows in post-ICV halos generally circularize before star formation.
As we discuss in Section \ref{sec:disc_lowz}, this explains why thin disk formation has been correlated with hot-mode inflows in post-ICV halos in FIRE.

\section{Results from the galaxy's edge to star formation}\label{sec:ResultsIII}
Finally, in this section we track gas particles after accretion onto the central galaxy.
We carry out galaxy-scale analyses of the kinematic properties of the gas before and after star formation, and the residence times of gas in galaxies.

\begin{figure*}
	\includegraphics[width=6.5in]{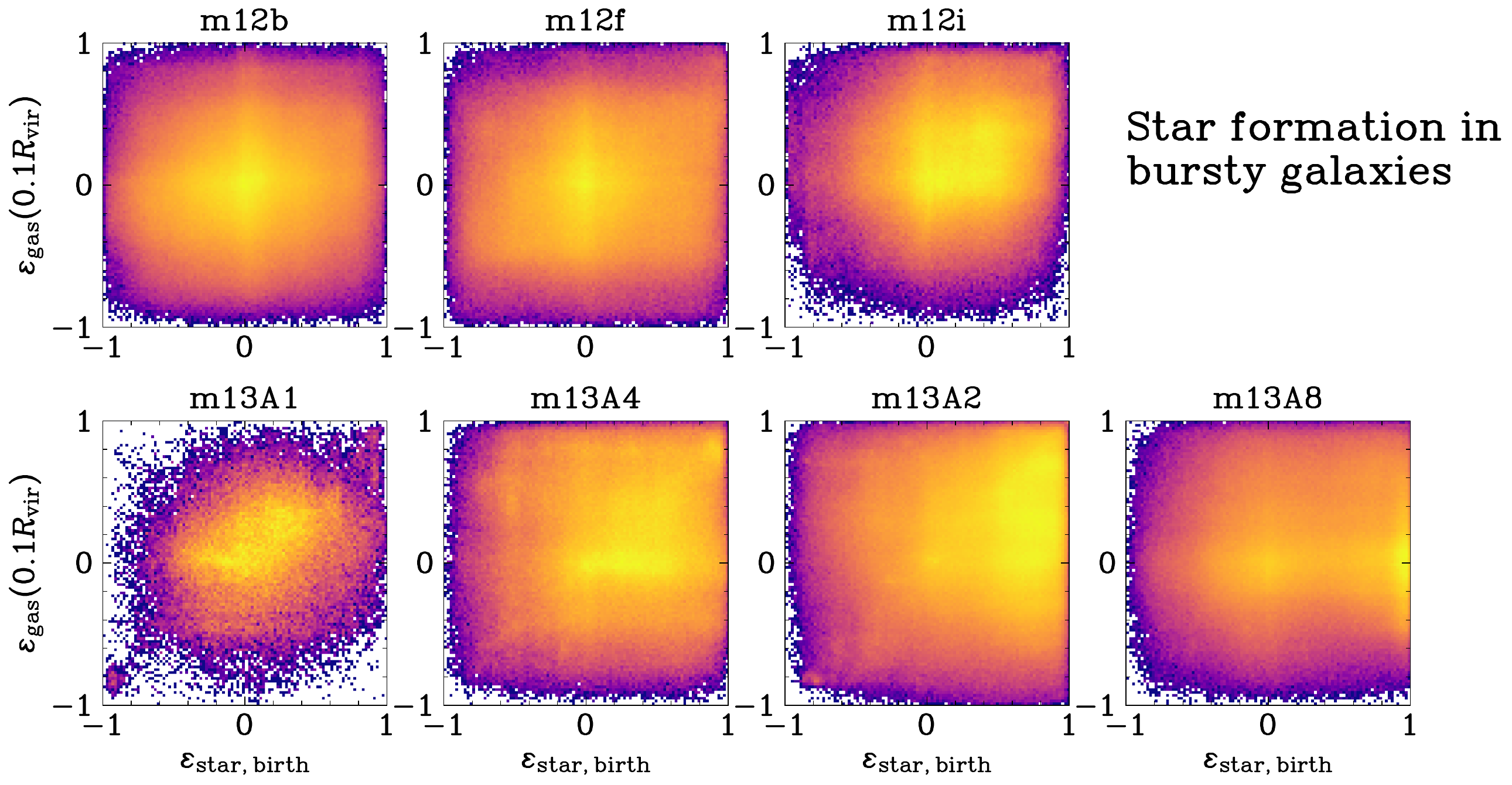}
    \caption{Comparison of the circularity of stars at birth to the circularity of their accreted gas progenitors, for stars that form in bursty galaxies with pre-ICV CGM.
    Only stars that formed \textit{in situ} at early times (i.e., stars that formed at $1.5<z<2.5$ for the m12, and from 500 Myr before $z_\mathrm{ICV}$ to $z=5$ for the m13 halos) are shown.
    $\epsilon_\mathrm{gas}(0.1R_\mathrm{vir})$ is the circularity of a gas particle at the last time it crosses $0.1R_\mathrm{vir}$ before flowing inward and forming a star with circularity $\epsilon_\mathrm{star,\ birth}$.
    The wide spread of circularities both at accretion and star formation reveal that accretion, dominated by cold gas in the bursty galaxies, does not circularize, and forms spheroid and thick disk stars.
    }
    \label{fig:epsilongas_earlytime}
\end{figure*}

\begin{figure*}
	\includegraphics[width=6.5in]{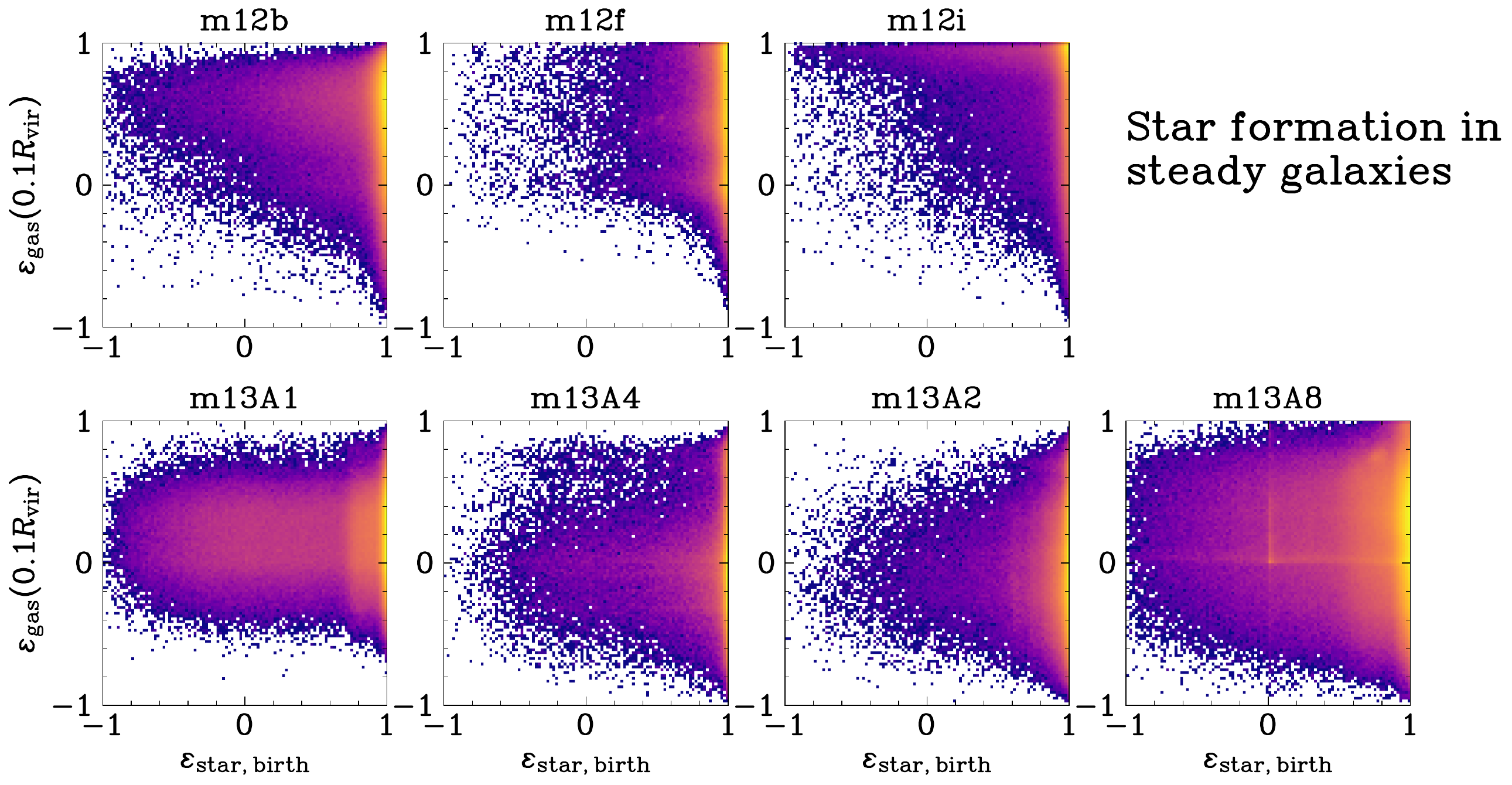}
    \caption{Comparison of the circularity of stars at birth to the circularity of their accreted gas progenitors, for stars that form in steady galaxies with virialized CGM.
    Similar to Figure \ref{fig:epsilongas_earlytime}, but \textit{in situ} stars that form at late times (i.e., during the last 1 Gyr of each simulation) are shown.
    Hot gas, which dominate the accretion in the steady galaxies, has not yet circularized at $0.1R_\mathrm{vir}$, but later circularizes and forms thin disk stars. 
    }
    \label{fig:epsilongas_latetime}
\end{figure*}

\begin{figure*}
	\includegraphics[width=6.5in]{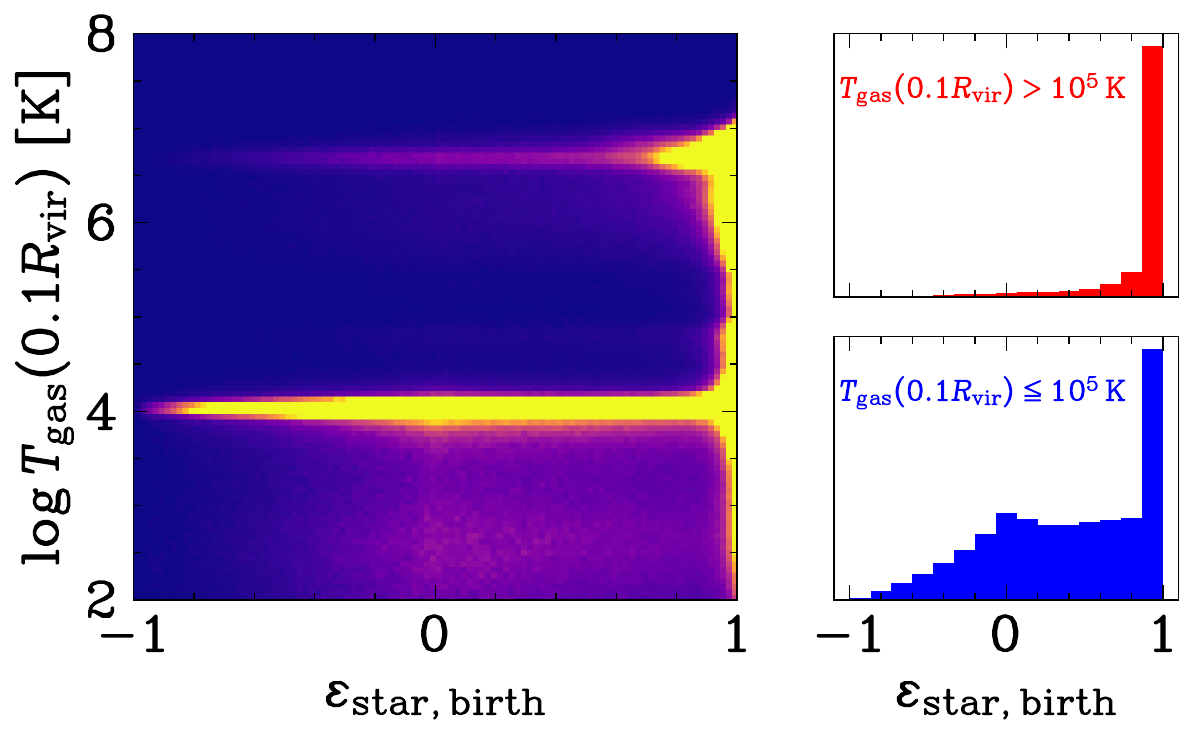}
    \caption{Comparison of stellar circularity at birth with the temperature of their progenitor gas at the time of accretion onto $0.1R_\mathrm{vir}$.
    The left panel shows the two-dimensional distribution stacked across seven galaxies in both low-mass bursty phases (Figure \ref{fig:epsilongas_earlytime}) and higher-mass steady phases (Figure \ref{fig:epsilongas_latetime}).
    The right panels show distributions of stellar circularity at formation for stars whose progenitor gas was hot (top) and cold (bottom) at the time of accretion onto $0.1R_\mathrm{vir}$.
    Stars with a broad range of circularities characteristic of spheroidal and thick-disk populations form from gas that is cold at the time of accretion, whereas nearly all of the stars that form from hot CGM gas are born with high circularities representative of thin-disk stellar populations. 
    A subdominant but non-negligible fraction of thin-disk stars form from cold progenitor gas.
    }
    \label{fig:Tgas_circ}
\end{figure*}

\subsection{Stellar particle tracking}\label{sec:results_startracking}
We track inflowing gas particles that eventually form stars in order to study their kinematic properties before and after star formation.
As described below, we identify the star particles residing in the central galaxy at the final simulation snapshot ($z=0$ for the m12 halos, and $z=1$ for the m13 halos) and trace them back to their progenitor gas particles.

\subsubsection{Back-tracking \textit{in situ} star formation}
We first select all star particles at $r<0.1 R_{\mathrm{vir}}$ at the final simulation snapshot.
Following \cite{yuBornThisWay2023}, we further restrict our selection to \textit{in situ} stars, i.e. stars that formed within $0.1 R_{\mathrm{vir}}$ of the most massive progenitor at their formation time.

For each star particle, we calculate the circularity of the particle $\epsilon$, which quantifies how the orbital angular momentum of the particle compares to that of a circular orbit with the same energy as the particle (see, e.g., \citealt{yuBurstyOriginMilky2021, yuBornThisWay2023}).
The circularity is defined as
\begin{equation}
    \epsilon(E) = \frac{\vec{j}\cdot\hat{z}}{j_\mathrm{circ}(E)},
\end{equation}
where $E=\frac{1}{2}v^2+\Phi(r)$ is the specific energy, $\vec{j}=\vec{v} \times \vec{r}$ is the specific angular momentum, and $\vec{v}$ and $\vec{r}$ are the velocity and position of the particle in the reference frame of the galaxy.
Additionally, $\Phi(r)$ is the gravitational potential at the radius $r \equiv |\vec{r}|$ of the particle, which we define as $\Phi(r) = \int^r_{3R_\mathrm{vir}} GM(<r)/r^2 \dd{r}$, where the enclosed mass of dark matter, gas, and stars, $M(<r)$, is calculated in logarithmically spaced radial shells.
Finally, $j_\mathrm{circ}(E) = v_c(r_E)r_E$ is the specific angular momentum of a circular orbit with radius $r_E$ that has the same specific energy as the particle.

For each star particle tracked, we measure the circularity at the formation time of the star particle, $\epsilon_\mathrm{birth}$.
As done by \cite{yuBornThisWay2023}, the stars can then be separated into ``spheroid'' ($\epsilon <0.2$), ``thick disk'' ($0.2<\epsilon <0.8$), and ``thin disk'' ($\epsilon>0.8$) classes based on their circularity at birth.
Note that this is a kinematic classification, while the spheroid, thick disk, and thin disk components of galaxies are often identified in observational studies of galaxy morphology (e.g., \citealt{juricMilkyWayTomography2008}).

Additionally, for each star particle that forms, we link the star particle to the gas particle out of which it formed.
We then track the properties of the \textit{progenitor} gas particle, including temperature, position, and circularity, back in time.

\subsubsection{Stellar and progenitor gas circularities}\label{sec:results_accretioncircularities}

We examine the relationship between the kinematic properties of stars and the gas from which they formed.
Our goal is to test whether star particles retain memory of their properties at the time of accretion.
We consider stars that form both in bursty and steady galaxies. 

Two-dimensional histograms are shown in Figure \ref{fig:epsilongas_earlytime} comparing the circularity of early-forming stars at birth to that of their accreted gas progenitors.
We include the stars that formed \textit{in situ} at $1.5<z<2.5$ for the low-redshift MW-mass halos, and from 500 Myr before ICV to $z=5$ for the high-redshift more massive halos.
The galaxies are bursty and their inner CGM have not virialized in these redshift ranges.
The circularity of a progenitor gas particle at accretion, $\epsilon_\mathrm{gas}(0.1R_\mathrm{vir})$, is the circularity of the gas particle at the last time it crosses $0.1R_\mathrm{vir}$ before flowing inward and forming a star of circularity $\epsilon_\mathrm{star,\ birth}$.

For the seven halos we analyze, there is a wide spread in both $\epsilon_\mathrm{gas}(0.1R_\mathrm{vir})$ and $\epsilon_\mathrm{star,\ birth}$ for the stars that formed at early times.
The large scatter in the circularities of stars at birth around $\epsilon_\mathrm{star,\ birth}\sim0$ that is the case for most of the halos is expected since star formation at early times is dominated by spheroid and thick disk populations with low circularities (see Appendix \ref{app:tformtICV} and \citealt{yuBornThisWay2023}).

In Figure \ref{fig:epsilongas_latetime}, we show the results above for stars that form during the last 1 Gyr of each simulation.
For stars that form at late times, there tends to be a wide spread in $\epsilon_\mathrm{gas}(0.1R_\mathrm{vir})$, while $\epsilon_\mathrm{star,\ birth}$ typically has very high values of $\sim1$.
This is consistent across the simulations.
As found by \citealt{yuBornThisWay2023, sternVirializationInnerCGM2021, gurvich_rapid_2023}, star formation in galaxies with steady star formation rates is dominated by the thin disk population (see also Appendix \ref{app:tformtICV}).

Figures \ref{fig:epsilongas_earlytime} and \ref{fig:epsilongas_latetime} show that gas which forms stars in both bursty and steady galaxies has a broad distribution of circularities at $0.1R_\mathrm{vir}$.
The circularities at accretion onto $0.1R_\mathrm{vir}$ do not show a correlation with the circularities at star formation, indicating that star particles do not retain memory of the kinematic state of the gas at the time of accretion.
In bursty galaxies, this may be due to the highly turbulent ISM, which broadly scatters stellar circularities.
In the steady galaxies, this may instead reflect the long timescales that accreted gas resides in the ISM prior to star formation and after it circularizes at $r<0.1R_\mathrm{vir}$.
During this `residence' time period, hydrostatic interactions within the ISM may effectively erase the kinematic record of the gas at the time of its accretion.
In the next section, we measure the residence times of accreted gas particles in bursty and steady galaxies prior to star formation, and compare them to the dynamical timescales of the galaxies.

\subsubsection{Gas Accretion Temperature and Stellar Circularity}\label{sec:results_accretiontemperatures}
We additionally investigate how the thermal state of gas at the time of accretion relates to the circularity of stars that form from it.
In the left panel of Figure \ref{fig:Tgas_circ}, we show the distribution of gas temperature at accretion, $T_\mathrm{gas}(0.1R_\mathrm{vir})$ as a function of stellar circularity at the time of star formation $\epsilon_\mathrm{star,\ birth}$.
$T_\mathrm{gas}(0.1R_\mathrm{vir})$ is defined as the circularity of a gas particle at the last time it crosses $0.1R_\mathrm{vir}$ before it forms stars.
The right panels in Figure \ref{fig:Tgas_circ} show distributions of $\epsilon_\mathrm{star,\ birth}$ for hot and cold progenitor gas.

For the results in Figure \ref{fig:Tgas_circ}, we stack two-dimensional distributions for \textit{in situ} stars forming in the seven galaxies in our analysis set in both the low-mass bursty time periods shown in Figure \ref{fig:epsilongas_earlytime}, and in the higher-mass steady periods shown in Figure \ref{fig:epsilongas_latetime}.
Each galaxy contributes one histogram per phase, and all 14 histograms are combined with equal weight.

Stars with a broad range of circularities indicative of spheroidal and thick-disk populations form out of cold $T \le10^5$ K CGM gas.
In contrast, nearly all of the stars that form out of hot $T > 10^5$ K CGM gas are born with circularities near $\epsilon_\mathrm{star,\ birth} \sim 1$, indicative of thin disk populations.
The $\epsilon_\mathrm{star,\ birth}$ distribution for cold progenitor gas also has a noticeable peak near $\sim 1$, indicating that some fraction of thin disk stars form from cold progenitor gas.
This is consistent with our results in Section \ref{sec:results_Mdot} where we found that inflows in the halos of steady galaxies (in which thin disk stars dominate the star formation) are predominantly hot but can contain small amounts of cold gas, especially at inner CGM scales (see also the temperature maps in Figures \ref{fig:Tmaps_halo} and \ref{fig:Tmaps}).
Previous FIRE studies have also found cool gas in the inner CGM of hot halos, which may originate from hot gas that cools as a result of nonlinear thermal instabilities (e.g., see  \citealt{esmerianThermalInstabilityCGM2021, sultanCoolingFlowsReference2025}).

\subsection{Residence time of gas in galaxies}\label{sec:results_residencetimes}

\begin{figure*}
	\includegraphics[width=6.5in]{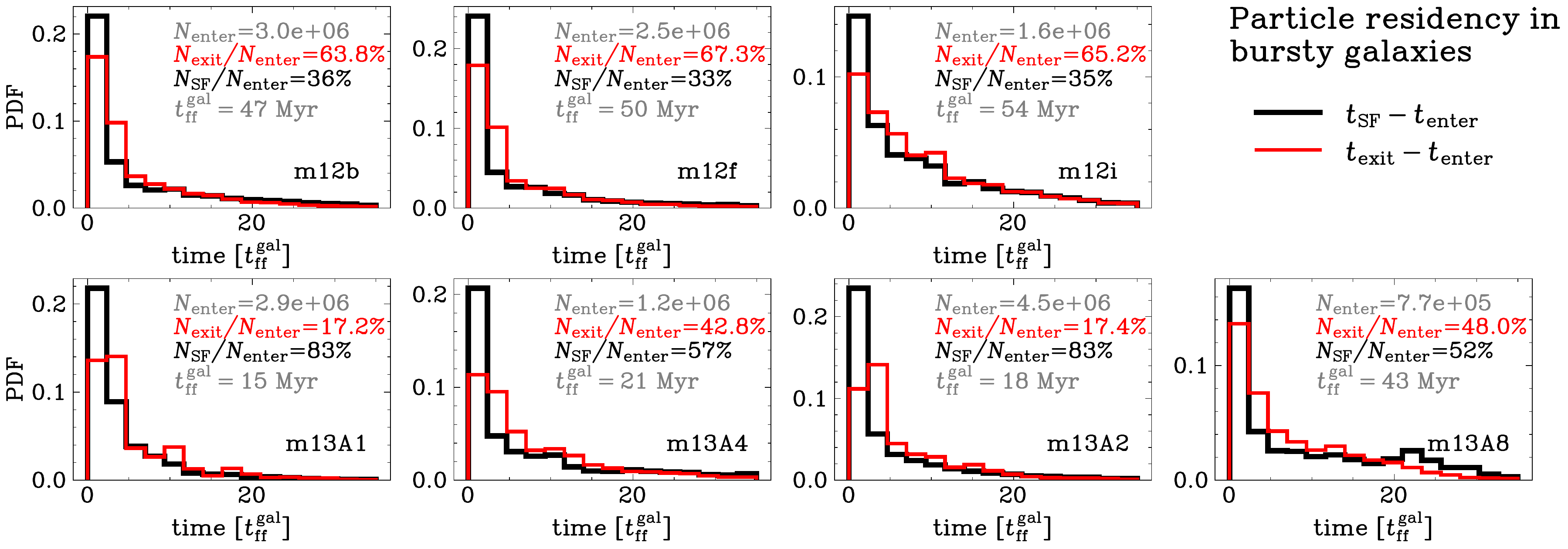}
    \caption{Distributions of the time gas accreted onto $r<0.05R_\mathrm{vir}$ spends in bursty galaxies before forming stars ($t_\mathrm{SF}-t_\mathrm{enter}$) or being ejected from the ISM ($t_\mathrm{exit}-t_\mathrm{enter}$).
    Gas accreted during a 1 Gyr time interval when galaxies are bursty (starting at $z=2.5$ for m12 halos and $z=5$ for m13 halos) is tracked to the end of the simulation. 
    For the $t_\mathrm{exit}-t_\mathrm{enter}$ distributions, we include only particles that are in strong outflows, i.e., particles that have enough energy to reach $0.5R_\mathrm{vir}$ after leaving the galaxy.
    Timescales are normalized by the galaxy free-fall time at $0.05R_\mathrm{vir}$ at the time of accretion for each particle, and the galaxy free-fall time at the midpoint of the accretion interval is listed in each panel for reference.
    Each panel also lists the total number of accreted particles during the 1 Gyr interval, as well as the fractions that form stars or are ejected as outflows during the tracking period.
    In highly dynamic, bursty galaxies, star-forming gas typically resides in the galaxy for $\lesssim 5 t_\mathrm{ff}^\mathrm{gal}$ before star formation.
    }
\label{fig:residencetimes_earlytime}
\end{figure*}

\begin{figure*}
	\includegraphics[width=6.5in]{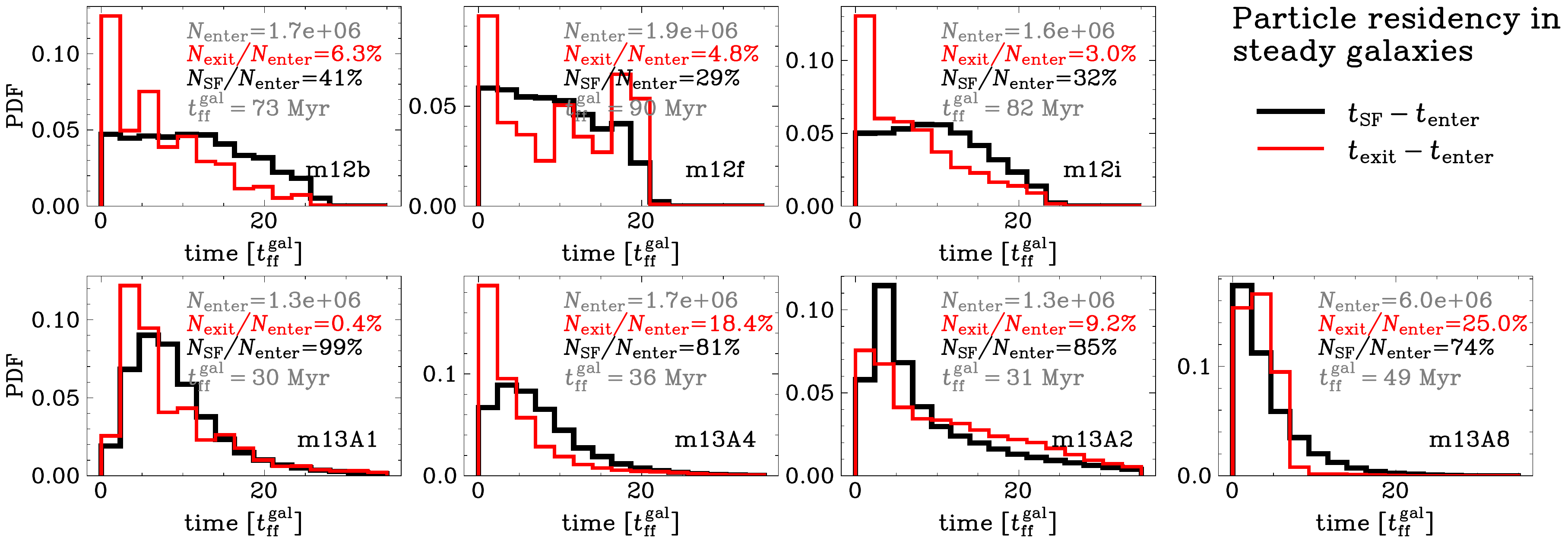}
    \caption{Distributions of the time gas accreted onto $r<0.05R_\mathrm{vir}$ spends in steady galaxies before forming stars ($t_\mathrm{SF}-t_\mathrm{enter}$) or being ejected from the ISM ($t_\mathrm{exit}-t_\mathrm{enter}$).
    Similar to Figure \ref{fig:residencetimes_earlytime}, but gas accreted 1-2 Gyr before the end of each simulation is tracked to the end of the simulation.
    In steady galaxies, accreted gas persists for up to $\sim25 t_\mathrm{ff}^\mathrm{gal}$ before collapsing into stars.
    }
\label{fig:residencetimes_latetime}
\end{figure*}

Once gas is accreted onto a galaxy, it can have one of two fates: it can either collapse and form stars, or be ejected from the ISM in an outflow prior to star formation.
In this section, we measure the timescales associated with the two possible outcomes, and compare them with the dynamical timescale of the galaxy.

We start by tracking all gas particles that cross the galaxy radius, which we approximate as $R_\mathrm{gal}=0.05R_\mathrm{vir}$, from outer radii during a given tracking window.
As described in Section \ref{sec:methods_halocentering}, we use $0.05 R_{\mathrm{vir}}$ as an approximation for the outer boundary of a galaxy rather than, e.g., measures based on the stellar half-mass radius $R_{s,1/2}$.
This choice provides a more stable reference radius and applies equally well to both bursty and steady galaxies.
In contrast, $R_{s,1/2}$ is prone to short-timescale fluctuations, particularly in low-mass galaxies undergoing bursty star formation and feedback cycles. 
For comparison, at late times our m12 galaxies form 50\% of their new stars within $\sim 0.02-0.03 R_{\mathrm{vir}}$, while most of the total stellar mass is contained within $\sim 0.01 R_{\mathrm{vir}}$ \citep[][]{gurvich_rapid_2023}. 
Eventually, the m13 galaxies in the steady regime become overly compact with stellar half-mass radii as small $\mathrm{few} \times 10^{-3} R_{\mathrm{vir}}$, which we interpret as a failure due to the neglect of AGN feedback, as described in Section \ref{sec:FIREmethod}.

When a gas particle that started outside the galaxy inflows to $r<0.05R_\mathrm{vir}$, we designate the time associated with the snapshot as $t_\mathrm{enter}$.
If the particle collapses into a star by the end of the simulation ($z=0$ for m12 halos, and $z=1$ for m13 halos) while still within the galaxy ($r<0.05R_\mathrm{vir}$), we designate the time of star formation $t_\mathrm{SF}$.
The amount of time between accretion and star formation is then $t_\mathrm{SF}-t_\mathrm{enter}$.

On the other hand, if the gas particle exits the galaxy ($r>0.05R_\mathrm{vir}$) before the end of the simulation while still a gas particle, we assign an exit time $t_\mathrm{exit}$.
The residence time of the particle, i.e. the amount of time between accretion and ejection, is then $t_\mathrm{exit}-t_\mathrm{enter}$.
For this, we apply additional outflow criteria to exclude particles whose radii fluctuate around $r=0.05R_\mathrm{vir}$ due to turbulent motion.
Specifically, we include only particles that satisfy the outflow criteria outlined in \cite{pandyaCharacterizingMassMomentum2021} (see their Section 3.1.1) at $t_\mathrm{exit}$.
Briefly, they select outflowing particles as particles for which 
\begin{align}
    &v_r>0\\
    &v_\mathrm{B,total}^2(r) > -\frac{1}{2}v_\mathrm{esc}^2(r_2),
\end{align}
where $v_\mathrm{B,total}$ is the total Bernoulli velocity, given by
\begin{equation}
    v_\mathrm{B,total}^2 = \frac{1}{2}v_r^2 + \frac{c_s^2}{\gamma-1} - \frac{1}{2}v_\mathrm{esc}^2.
\end{equation}
$c_s=\sqrt{\gamma \frac{k_\mathrm{B} T}{\mu m_\mathrm{p}}}$ is the sound speed, where the adiabatic index $\gamma=5/3$ for a monatomic ideal gas, and $v_\mathrm{esc}$ is the escape velocity.
These outflow criteria select outflowing particles that have enough energy to reach radius $r_2$ from their current radius $r$.
We choose an outer radius of $r_2 = 0.5R_\mathrm{vir}$.

If an accreted gas particle reaches $r>0.05R_\mathrm{vir}$ without satisfying the outflow criteria defined above, we continue to track the particle.
If it remains a gas particle and satisfies the outflow criteria at a later time, that later time is defined as $t_\mathrm{exit}$ and the particle is included in the gas residence-time distribution, $t_\mathrm{exit}-t_\mathrm{enter}$.
If it instead forms a star, either after re-entering $0.05R_\mathrm{vir}$ or outside this radius, we include it in the star formation time distribution, $t_\mathrm{SF}-t_\mathrm{enter}$. 
Accreted particles that reach $r>0.05R_\mathrm{vir}$, remain gas until the final snapshot, and never satisfy the outflow criteria are not included in either the stellar or gas residence-time distributions.
These particles are the primary reason that the fractions of accreted particles that either form stars or are ejected from the galaxy in outflows, indicated in Figures~\ref{fig:residencetimes_earlytime} and \ref{fig:residencetimes_latetime}, do not sum exactly to unity.

In Figures \ref{fig:residencetimes_earlytime} and \ref{fig:residencetimes_latetime}, we present distributions of $t_\mathrm{SF}-t_\mathrm{enter}$ and $t_\mathrm{exit}-t_\mathrm{enter}$ for bursty and steady galaxies, respectively.
Results for gas that is accreted during a 1 Gyr time interval when galaxies are bursty (starting at $z=2.5$ for the m12 halos and $z=5$ for m13 halos) and is tracked to the end of the simulation are shown in Figure \ref{fig:residencetimes_earlytime}.
Figure \ref{fig:residencetimes_latetime} shows results for gas accreted 1-2 Gyr before the end of each simulation ($0.08<z<0.16$ for the m12 halos, and $1.3<z<1.7$ for the m13 halos) and tracked to the end of the simulation.
We show the time distributions in units of the galaxy-scale free-fall time at the time a particle is accreted onto $0.05 R_\mathrm{vir}$, defined as
\begin{equation}
    t_\mathrm{ff}^\mathrm{gal}=\frac{\sqrt{2}R_\mathrm{gal}}{v_c(R_\mathrm{gal})},
\end{equation}
with $R_\mathrm{gal}=0.05R_\mathrm{vir}$.

In bursty galaxies, Figure \ref{fig:residencetimes_earlytime} reveals that most gas resides in the galaxy for a short amount of time ($\lesssim 5-10\,t_\mathrm{ff}^\mathrm{gal}$) before either forming stars or exiting, with both populations of particles following roughly the same distributions.
As indicated in the panels, $\sim30-80$\% of gas accreted during the bursty phase forms stars.
Large fractions of accreted gas are ejected from the galaxies ($\sim20-70$\%) in strong outflows by the end of the simulations.

In steady galaxies, Figure \ref{fig:residencetimes_latetime} shows that gas can reside in the ISM for tens of galaxy free-fall times before forming stars. 
The probability distributions of $t_\mathrm{SF}-t_\mathrm{enter}$ are generally broader and gas has a higher probability of residing in galaxies for up to $\sim20-25 t_\mathrm{ff}^\mathrm{gal}$ before star formation.
As discussed in Section \ref{sec:results_accretioncircularities}, long residence times prior to star formation allow hydrostatic forces to wipe the kinematic record of the gas at the time of its accretion, which may explain the lack of correlation apparent in Figure \ref{fig:epsilongas_latetime} between the circularities of a particle at accretion and star formation.

In all of the simulations except m13A4 and m13A8, $\lesssim10$\% of accreted gas in steady galaxies exits while satisfying the outflow criteria.
This is consistent with previous FIRE studies which found that outflows from the ISM into the CGM are minimal at $z\sim0$ in MW-mass galaxies \citep{muratovGustyGaseousFlows2015, angles-alcazarCosmicBaryonCycle2017, pandyaCharacterizingMassMomentum2021}.
The higher exit fractions in m13A4 and m13A8 of $\sim20$\% suggests the galaxy might have a significant outflow for some time during the steady phase. 
Note that significant outflows in galaxies that have entered their ``steady'' phase can occur when they experience short-lived starbursts, such as may be triggered by galaxy mergers. 
In some cases, particularly for the m12 halos, the fractions of accreted particles that form stars or are ejected as gas in Figure \ref{fig:residencetimes_latetime} sum to significantly below 100\%.
This reflects both particles that remain in the galaxy without sufficient time to form stars or exit by the end of the simulation, as well as particles that leave the galaxy in weak outflows not captured by our relatively stringent outflow criteria.
In particular, particles that do not have enough energy to reach $0.5R_\mathrm{vir}$ after leaving the galaxy are excluded from our analysis.

Figures \ref{fig:residencetimes_earlytime} and \ref{fig:residencetimes_latetime} demonstrate that at early times in galaxy formation, star-forming gas generally resides in galaxies for $\lesssim 5 t_\mathrm{ff}^\mathrm{gal}$ before star formation, while at late times, gas is able to persist up to $\sim25 t_\mathrm{ff}^\mathrm{gal}$ before collapsing into stars.
The galaxies in the early time windows, as defined here, are hosted by halos that have not completed virialization.
As we discussed in Section \ref{sec:burstytosteadySFR}, the pre-ICV galaxies are highly dynamic and have strong bursts of star formation and subsequent stellar feedback (e.g., \citealt{muratovGustyGaseousFlows2015, sternVirializationInnerCGM2021, gurvich_rapid_2023}).
This is in sharp contrast with the galaxies in the late time windows, which are post-ICV galaxies with steady star formation.
As we discuss in Section \ref{sec:disc_burstysteady}, this analysis points to a key difference in how star formation is regulated in bursty galaxies, shown by \cite{sunTurbulentFrameworkStar2026} to be well modeled by a CGM-scale turbulent framework, as opposed to steady equilibrium disks.

\section{Discussion}\label{sec:Discussion}

In this section, we discuss our results in the context of previous studies. 

\subsection{Angular momentum in the halo from cold and hot accretion}\label{sec:disc_angmom}
Our halo-scale analysis of the CGM in FIRE (Section \ref{sec:ResultsI}) broadly agrees with the results of past studies exploring a wide array of simulations, which have found that gas in halos contains greater amounts of sAM than dark matter, and cold CGM gas is primarily responsible for this effect (e.g., \citealt{danovichFourPhasesAngularmomentum2015, stewartHighAngularMomentum2017, defelippisAngularMomentumCircumgalactic2020, wangLargescaleEnvironmentCGM2021}).
In our FIRE halos over the redshift we probed, cold inflows dominate the sAM of the halos.

\citet{danovichFourPhasesAngularmomentum2015} linked the high sAM of cold inflows to their thin stream-like geometry.
They found that tidal torques impart angular momentum to streams as they approach the halo more efficiently than to the more isotropic dark matter.
Once the cold streams enter the halo, they flow to the inner halo with minimal changes to their sAM, while the dark matter and hot gas are more prone to mixing, reflected by their much lower sAM coherence that \cite{danovichFourPhasesAngularmomentum2015} measured.
We similarly found that the cold inflows in FIRE are more coherent in their angular momentum than the hot inflows at halo-scale radii.

In the next section, we discuss the structure of the cold gas in our virialized halos, particularly the m13 halos which virialize at high redshift ($z_\mathrm{ICV}\gtrsim1.9$).
Cold inflows are present in the m13 halos at the time of ICV, but the cold gas is morphologically complex and much more disordered than the narrow cold streams predicted by analytic models.

\subsection{Reassessing the significance of cold mode accretion in the inner CGM of massive galaxies at high redshift}\label{sec:disc_pastwork}

As summarized in our introduction, previous studies predicted the existence of cold streams originating from cosmic web filaments that penetrate the hot gas in virialized massive halos at high redshifts.
\cite{dekelColdStreamsEarly2009} proposed a picture of high-redshift massive disk formation in which cold streams dominated the accretion, rather than mergers, and were responsible for producing the high measured star formation rates in galaxies at $z\sim2.5$.

This idea was motivated in part by work by \cite{dekelGalaxyBimodalityDue2006} on the bimodality in galaxy properties found in many observational studies, where massive halos host red, quiescent elliptical galaxies, while blue, star-forming disk galaxies dominate galaxy populations at lower masses, with the evolution in galaxy properties occurring at a critical halo mass of $\sim10^{12}~\Msun$.
To explain the origin of the bimodality, \cite{dekelGalaxyBimodalityDue2006} used a shock stability criterion to derive an analytic prediction of the critical halo mass marking the transition from cold to hot halos, building on earlier results by \cite{birnboimVirialShocksGalactic2003}. 
They showed that the metallicity of the accreted gas and the radius of the accretion shock, which determine the critical mass in their model, can be chosen such that the critical virialization mass is approximately the critical mass of the galaxy bimodality $\sim10^{12}~\Msun$, suggesting that star-forming spiral galaxies are fed by cold accretion.

The m13 halos we analyze finish virializing at high redshift ($z_\mathrm{ICV}\sim1.9-3.6$), during the epoch when cold streams are suggested to penetrate virialized halos.
As we show in the second row of Figure \ref{fig:Mdotin_allz}, the fractions of hot and cold inflowing gas in the m13 halos are comparable at the time the inner halo virializes.
While there is a substantial cold inflow fraction at the time of ICV, we find hot gas increasingly dominates the inflow as the halo evolves after ICV, in contrast with the picture of \cite{dekelColdStreamsEarly2009}, in which cold streams dominate the accretion.

In our edge-on temperature maps, at post-ICV times there is no significant filamentary feeding from large scales that persists down to the central galaxy (Figures \ref{fig:Tmaps_halo} and \ref{fig:Tmaps}).
It is possible the cold streams do not survive very deep into the halo.
Cold stream penetration depths vary widely across simulations, ranging from halo to galaxy scales (e.g., $\sim0.7-0.2\ R_\mathrm{vir}$ in \citealt{danovichCoplanarStreamsPancakes2012, nelsonZoomingAccretionStructure2016, medlockStatisticalPropertiesCold2025, waterval_gas_2025}).
As emphasized by \cite{medlockStatisticalPropertiesCold2025}, this large variation shows that analyses of cold streams depend strongly on the stream finding algorithm and the numerical methods used in the simulations.
In future work, it would be interesting to carry out a more careful identification of cold streams in our high-redshift FIRE halos from cosmological to galaxy scales.
Studies of idealized high-resolution hydrodynamical simulations have additionally found that the survival of cold streams can be affected by instabilities on small scales that are unresolved in our halos, while additional physics such as radiative cooling and a realistic halo potential may allow hot gas to cool and become entrained in the cold stream (e.g., \citealt{mandelkerInstabilitySupersonicCold2016, padnosInstabilitySupersonicCold2018, mandelkerInstabilitySupersonicCold2020, aungEntrainmentHotGas2024}).

To summarize, we find that in our high-redshift, virialized m13 halos, cold gas comprises $\sim50\%$ of the inflow at the time of ICV, but drops to only a few$-10\%$ by $z=1$.
In temperature maps, the cold gas in the virialized halos is morphologically complex and much more disordered than the steady ordered cold streams predicted by analytic models to dominantly feed the inner halo.
Rather, we find hot gas is responsible for most of the accretion in the inner CGM of virialized halos, at both low and high redshifts.
Next, we discuss the hot inflows and their role in disk formation.

\subsection{Hot rotating inflows dominate accretion onto galaxies in halos which form thin disks}\label{sec:disc_lowz}

A key finding by \cite{hafenHotmodeAccretionPhysics2022} is that in MW-mass FIRE halos at $z\sim0$, the emergence of thin galaxy disks is correlated with hot, rotating cooling flows dominating the accretion in the inner CGM.
The rotating cooling flows can be sustained in a virialized halo, as analytically described by \cite{sternAccretionDiscGalaxies2024}.
In this paper, we demonstrate that hot, rotating inflows are responsible for the bulk of the accretion in the inner CGM at both low redshifts (in hot halos) and high redshifts (in `cold-in-hot' halos). 
In all of the halos in our sample, we find that at times following ICV, hot gas increasingly dominates the inflow. 
The hot inflowing gas circularizes, cools below $10^5$ K, and forms stars, consistent with the $z\sim0$ analysis of \cite{hafenHotmodeAccretionPhysics2022}.
The stars in post-ICV halos are born with high circularities indicative of thin-disk populations, as opposed to the spheroid and thick disk stars formed in pre-ICV cold mode-dominated halos.
This is consistent with analyses of low-redshift MW-mass FIRE halos by \cite{yuBurstyOriginMilky2021, yuBornThisWay2023} and other previous FIRE results on the emergence of thin disks (e.g., \citealt{sternVirializationInnerCGM2021, gurvich_rapid_2023, myrtajProtogalaxyThickThin2026}).

The result in FIRE that stars formed at early times have a broad distribution of circularities characteristic of spheroid and thick disk populations, while thin disk stars dominate star formation at later times after ICV  (see Figure \ref{fig:starformationtimes}), is consistent with galactic archaeology studies pioneered by \cite{eggenEvidenceMotionsOld1962}.
\cite{eggenEvidenceMotionsOld1962} compared the kinematics of Milky Way stars to their metallicities, and, using metallicity as a proxy for stellar age, they found that older MW stars had dispersion-dominated kinematics, while younger stars had low eccentricities indicative of disk-like kinematics. 
They interpreted this discovery using a monolithic collapse picture, in which old stars form on highly elliptical orbits out of the collapsing protogalactic gas cloud, and rotationally-supported disk-like stars later form once the cloud collapses to a thin plane. 
Our simulations provide a more modern interpretation anchored in $\Lambda$CDM cosmology, %
where continuously growing low-mass galaxies repeatedly undergo bursts of star formation and subsequent feedback that inhibit long-lived rotationally-supported disks from forming, and galaxies only settle into thin disks once the star formation rate stabilizes at higher masses (e.g., \citealt{gurvich_rapid_2023}).
The phases of MW evolution uncovered by more recent archaeological studies such as \cite{belokurovDawnTillDisc2022} and \cite{chandraThreephaseEvolutionMilky2024}, namely a dispersion-dominated spheroid, a turbulent thick disk, and finally a dynamically-cold thin disk, are broadly consistent with stellar populations in FIRE cosmological simulations of MW-like galaxies (see \citealt{myrtajProtogalaxyThickThin2026} for a recent study).

In our picture, cold accretion in pre-ICV halos is associated with bursty galaxies that may be identified with forming thick disks, pseudobulges, or irregular galaxies.
Steady thin disks form at low redshift after ICV when hot gas dominates the inflow.
Our picture qualitatively agrees with many observational studies which find that large, thin, rotationally-supported disks at low redshift are common around the MW-mass scale, while lower mass galaxies have more irregular morphologies and dispersion-dominated kinematics (e.g., \citealt{simons2EpochDisk2017a, bizyaevSpectralObservationsSuperthin2021, tileyKMOSGalaxyEvolution2021}), although there appear to be interesting quantitative differences in the exact mass scale where thin disks become common in observations vs. in FIRE simulations \citep[e.g.,][]{el-badryGasKinematicsMorphology2018, 2025MNRAS.544.4651B, 2026ApJ...998..125K}.
Note we distinguish the irregular morphologies of low-mass FIRE galaxies from classical bulges \citep[as distinguished by e.g.][]{2004ARA&A..42..603K}, and caution against directly comparing our findings with, e.g., bulge-to-total stellar mass ratios.

A natural question that arises is why cold accretion carries the bulk of the angular momentum in the CGM across cosmic time, yet thin disks only emerge in FIRE once hot accretion dominates the inflow.
This contrasts with a common picture of disk formation in which cold streams are expected to dominate accretion onto galaxies and to promote disk formation by coherently delivering high amounts of specific angular momentum to the galaxy.
We find that, as in the low redshift halos studied by \cite{hafenHotmodeAccretionPhysics2022}, hot gas in both low- and high-redshift virialized halos circularizes before it cools and forms stars, while cold accretion does not circularize prior to star formation in pre-ICV halos where it dominates the inflow.
Thus, our results support a picture in which the circularization of inflows in the inner CGM prior to star formation, rather than the amount of specific angular momentum of gas in the halo, is correlated with the formation of thin disks.

As \cite{hafenHotmodeAccretionPhysics2022} discuss, a potential reason why hot rotating inflows are favorable for thin disk formation is that hot cooling flows are subsonic, with an inflow timescale that exceeds the sound-crossing time.
This allows enough time for hydrodynamical forces to align the angular momentum vectors of the inflowing gas prior to star formation.
Consistent with this picture, Figures \ref{fig:vphi_ICV} and \ref{fig:rscaled_ICV} show that the hot inflows circularize in the inner CGM immediately before cooling, implying synchronization of their angular momentum before accretion.
Cold inflows, on the other hand, reach the galaxy at supersonic velocities, and there is insufficient time for hydrodynamic forces to synchronize the angular momentum vectors of the cold gas before it forms stars.
Additionally, the hot inflows are smoother (with relatively modest density fluctuations) compared to cold inflows, in which gas is highly clumpy and has a density that fluctuates over many orders of magnitude (e.g., \citealt{kakolyTurbulencedominatedCGMOrigin2025, sunTurbulentFrameworkStar2026}).
The fragmented morphology of cold inflows may further inhibit them from settling into a thin coherently rotating disk. 

One notable difference we find in the high-redshift regime compared to lower redshifts is that gas disks in steady galaxies are thicker and have higher amounts of dispersion.
As shown in Figure \ref{fig:vphi_sigma}, during $\sim$2 Gyr immediately after $t_\mathrm{bursty}$, the m13 galaxies have $\left< v_\phi \right>/\sigma_g \lesssim 4$, while the m12 galaxies reach $\left< v_\phi \right>/\sigma_g \gtrsim 5$.
This systematic difference highlights an important point: although \cite{hafenHotmodeAccretionPhysics2022} showed that the hot accretion mode is linked to the formation of thin disks, our results demonstrate that hot accretion alone is not sufficient for thin disks to form.

In particular, the lack of thin disks at high redshifts suggests additional physics that drives turbulence in the ISM of high-redshift massive galaxies.
We find similar distributions of high circularities of stars at birth (indicative of thin disk populations) at both low and high redshift in steady galaxies, which hints that hot inflows circularize and form thin-disk stars in both regimes, but additional processes stir up turbulence in the high redshift regime, producing thicker disks.
Turbulence in the disks may be driven by processes such as the accretion of cold gas clumps, compressive tidal forces, and stellar feedback, although the relative impact of these effects is uncertain (e.g., \citealt{ginzburgEvolutionTurbulentGalactic2022, mandelkerFormationGiantClumps2025, ginzburgOriginCompressiveTurbulence2025}).
We note our sample of four high-redshift massive halos is small, and a larger suite of high-resolution simulations at high-redshift is required to test the robustness of our result.

Before concluding this section, we note that while the association of thin disk galaxies with hot mode accretion in FIRE (rather than cold streams) differs from some previous theories, it is consistent with some other previous theoretical predictions. 
For example, \cite{cattaneoGalICS21New2020} used a semi-analytic model expanding on \cite{dekelGalaxyBimodalityDue2006}'s results to predict the gas metallicity and accretion shock radius, and found a halo virialization mass of $\sim2\times10^{11}~\Msun$ at $z=0$, implying that at low redshift, %
disk galaxies like the Milky Way and M31, with $M_{\mathrm{vir}} \approx (1-2)\times 10^{12}$ M$_{\odot}$ \citep[e.g.,][]{2023MNRAS.521.4863S}, are hosted in virialized halos in which hot accretion dominates.

\subsection{Bursty vs. steady galaxies and their CGM}\label{sec:disc_burstysteady}

There are significant differences in our results bridging accretion and star formation between bursty and steady galaxies.
As discussed in Section \ref{sec:burstytosteadySFR}, many previous FIRE studies have identified a transition in galaxy properties from bursty star formation rates and strong outflows to steady star formation and weaker outflows, occurring at the $L^*$ mass scale (e.g., \citealt{sternVirializationInnerCGM2021, byrneStellarFeedbackregulatedBlack2023, muratovGustyGaseousFlows2015, angles-alcazarCosmicBaryonCycle2017}).
In both the m12 and m13 halos we analyze, we find that the inner CGM virializes at roughly the same time that the star formation rate of the galaxy stabilizes and the galaxy becomes less bursty.
This confirms and extends the result of \cite{gurvich_rapid_2023} for the evolution of MW-mass halos to more massive halos.
As \cite{hafenHotmodeAccretionPhysics2022} discuss, ICV might drive disk settling by ensuring that gas is accreted in a subsonic, coherently rotating cooling flow.

Once gas accretes onto the galaxy, we find it resides in post-ICV, steady galaxies for up to $\sim25$ galaxy free-fall times before forming stars.
On the other hand, gas typically resides in pre-ICV, bursty galaxies for less than $\sim5$ galaxy free-fall times before forming stars.
This difference reveals systematic differences in how star formation is regulated in bursty vs. steady galaxies.

Previous studies found that star formation in steady and bursty galaxies can be described by different analytic models.
For example, \cite{gurvichPressureBalanceMultiphase2020} showed that pressure balance in the ISM of steady FIRE galaxies at $z\sim0$ is consistent with `equilibrium disk' models, which describe a rotationally supported disk in which star formation obeys the observed Kennicutt-Schmidt relation (e.g., \citealt{thompsonRadiationPressureSupported2005, ostrikerMAXIMALLYSTARFORMINGGALACTIC2011, faucher-giguereFeedbackregulatedStarFormation2013}).
Recently, \cite{sunTurbulentFrameworkStar2026} showed that star formation in bursty FIRE galaxies at high redshifts could be modeled as occurring in high density regions of a supersonically turbulent gas medium that extends from ISM to CGM scales.
They noted that their turbulent model is likely applicable to bursty galaxies at lower redshifts as well because the essential assumption is that the inner CGM is in the rapid cooling limit ($t_\mathrm{cool}^\mathrm{(s)} \lesssim t_\mathrm{ff}$), which is expected in low-mass halos at all redshifts.

Our analyses of gas accretion relative to star formation times (Section \ref{sec:results_parttracking_tform}) and of residence times (Section \ref{sec:results_residencetimes}) have implications for how gas accretes and forms stars in the two regimes.
In the turbulent-halo star formation scenario in bursty galaxies, cold gas does not generally circularize, and forms stars after residing in the galaxy for only a few galaxy free-fall times.
On the other hand, hot gas circularizes as it accretes onto the equilibrium disk of steady galaxies, and forms stars on significantly longer time scales because only modest star formation is needed for stellar feedback to support the disk against gravity.

\subsection{Comparison to Observations of Disk Kinematics}
Finally, we briefly discuss our results in the context of observations of disk formation, beginning with measurements of the fraction of star-forming galaxies that have disks across redshift.
A rigorous comparison of the velocity dispersions we measure in our simulations with observations requires a careful modeling of observables (e.g., kinematics traced by ionized vs. molecular gas) that is beyond the scope of this paper.

\cite{tileyKMOSGalaxyEvolution2021} studied the kinematics of $>200$ star-forming galaxies of stellar masses $\sim10^9-10^{11.5}\ \Msun$ at $z\sim1.5$ as part of the KMOS Galaxy Evolution Survey (KGES).
They observed H $\alpha$ and $\left[\mathrm{N\ II}\right]$ emission and calculated ionized gas rotational velocities and velocity dispersions, which they compared with hundreds of star-forming galaxies observed by the KROSS ($z\sim0.9$) and SAMI ($z\sim0$) surveys.
They included galaxies from the three surveys with spatially resolved H $\alpha$ emission in their comparison sample.
For the galaxies in their comparison, \cite{tileyKMOSGalaxyEvolution2021} measured disk fractions, which they defined as the fraction of galaxies with $v_\phi/\sigma_g$ above a given threshold and whose ISM velocity field was well fit by an analytic disk model.
Across $0\lesssim z \lesssim 1.5$, they found that disk fractions strongly depend on the stellar mass, and have a much weaker dependence than redshift.
At fixed redshift, disk fractions found by \cite{tileyKMOSGalaxyEvolution2021} vary by up to $\sim50\%$ between their three stellar mass bins.
This finding broadly agrees with our result in Figure \ref{fig:vphi_sigma} that the FIRE galaxies in our analysis have a transition in their star formation rates from bursty to steady that correlates with an evolution from dispersion-dominated to rotationally-supported kinematics in their ISM.
The bursty-to-steady SFR transition at $t_\mathrm{bursty}$ occurs at the $\sim L^*$ galaxy mass regardless of redshift, above which we expect most star-forming galaxies to have disks (see also, \citealt{gurvich_rapid_2023}).

Additionally, we find that the high-redshift steady galaxies in our FIRE analysis are dynamically hotter (i.e., have a lower $v_\phi/\sigma_g$) than the low-redshift steady galaxies (Figure \ref{fig:vphi_sigma}), which is due to greater velocity dispersions in the m13 galaxies.
\cite{johnsonKMOSRedshiftOne2018} analyzed integral field unit data from the KROSS ($z\sim0.9$) and SAMI ($z\sim0$) surveys, and MUSE observations of $\left[ \mathrm{O\ II}\right]$ emission in $\sim100$ galaxies at $z \sim 0.5$.
The galaxies in their sample span stellar masses $\sim10^8-10^{11}\ \Msun$, and they find $\sim50\%$ higher average velocity dispersions at $z\sim1$ than at $z\sim0$ at fixed stellar mass.
Other observational studies of ionized gas have similarly measured higher velocity dispersions in high redshift disks than low redshift disks at fixed stellar mass (e.g., \citealt{wisnioskiKMOS3DSURVEYDESIGN2015, simons2EpochDisk2017, priceMOSDEFSurveyKinematic2020}).
The m12 galaxies in our analysis reach similar (though somewhat lower) stellar masses at $z=0$ to the m13 halos at $z=1$ ($M_*\sim10^{11}\ \Msun$), and the higher velocity dispersions we measure in the m13 galaxies at $z=1$ relative to the m12 galaxies at low redshift are in qualitative agreement with the observational trend.

\section{Conclusions}\label{sec:Conclusions}
In this paper, we systematically study gas accretion and angular momentum in high resolution FIRE cosmological zoom-in simulations of galaxies.
Our sample consists of seven halos covering a wide range in mass ($M_{\mathrm{halo}} \sim 10^{10.5}-10^{13}\ \Msun$) and redshift ($z=5$ to $z=0$), including high-redshift halos in the `cold-in-hot' regime, where cold streams are expected to penetrate a hot, virialized CGM. 

Using a combination of instantaneous measurements of inflows in the halo and galaxy properties, and particle tracking of gas accretion onto the galaxy, we investigate gas accretion, angular momentum delivery, star formation, and disk formation.
We perform our analysis at three physical scales, from the scale of the halo ($\sim R_\mathrm{vir}$) down to the scale of the galaxy ($\lesssim 0.05R_\mathrm{vir}$).

\textbf{On halo scales}, we find that:
\begin{enumerate}
    \item There is an evolution in the mode of gas accretion onto galaxies which correlates with the virialization of the inner CGM.
    Inflows are almost entirely cold ($T<10^5$ K) at times $\gtrsim$ Gyr prior to ICV, while the hot phase ($T>10^5$ K) dominates the inflow at times $\gtrsim$ Gyr after virialization of the inner halo (Figure \ref{fig:Mdotin_allz}).
    Massive galaxies have hot-mode dominated accretion, including high redshift galaxies in the `cold-in-hot' regime for which cold streams have been predicted by previous works.
    
    \item Cold inflows carry more specific angular momentum than both dark matter and hot inflows at $r\gtrsim 0.1 R_\mathrm{vir}$, consistent with previous simulation studies (Figure \ref{fig:jtot_allz}).
\end{enumerate}

\textbf{On the scale of the inner CGM and during accretion onto the galaxy}, we find that:
\begin{enumerate}
    \item Hot inflowing gas rapidly circularizes in the inner CGM of virialized halos simultaneously with cooling below $10^5$ K (Figures \ref{fig:vphi_ICV} and \ref{fig:rscaled_ICV}). 
    This behavior is consistent across both high- and low-redshift post-ICV halos.

    \item Whether inflowing gas circularizes before star formation strongly depends on the state of the CGM.
    In bursty low-mass galaxies, the cold inflowing gas that dominates accretion in the inner CGM does not circularize prior to star formation (Figure \ref{fig:vphi_tstarform_earlytimes}).
    In steady massive galaxies with a virialized halo, the hot inflowing gas that dominates accretion in the CGM tends to rapidly circularize shortly before star formation (Figure \ref{fig:vphi_tstarform_latetimes}).
\end{enumerate}
The two results above provide an explanation for why the formation of thin disks has been linked with hot mode accretion (e.g., \citealt{hafenHotmodeAccretionPhysics2022}), despite cold gas carrying a greater amount of specific angular momentum in the halo.

Finally, \textbf{on galaxy scales following accretion}, we find that:
\begin{enumerate}
    \item Consistent with previous studies of FIRE simulations, stars that form well before ICV are born with a broad distribution of circularities (indicative of spheroidal and thick-disk populations), while stars formed after ICV have high circularities characteristic of thin-disk populations.
    We find no correlation between the circularity of gas at accretion onto $0.1R_\mathrm{vir}$ and the circularity of the stars it subsequently forms (Figures \ref{fig:epsilongas_earlytime} and \ref{fig:epsilongas_latetime}).
    In pre-ICV halos, this may reflect the highly turbulent ISM of the bursty galaxy, which broadly scatters stellar circularities, while in post-ICV halos, gas may reside in the galaxy long enough after circularization for hydrodynamic forces to erase its kinematic memory before forming stars.

    \item Stars with a broad range of circularities (representative of spheroidal and thick-disk populations) form out of gas that is cold at the time of accretion onto $0.1R_\mathrm{vir}$, while the vast majority of stars that form out of hot CGM gas are born with high circularities corresponding to thin-disk stellar populations.
    A subdominant but non-negligible fraction of thin disk stars form from cold progenitor gas (Figure \ref{fig:Tgas_circ}).

    \item We compare disk formation in FIRE in high and low redshift MW-mass galaxies. 
    While transitions in galaxy properties (e.g., the transition from bursty to steady star formation rates shown in Figure \ref{fig:SFHscatter}, and the transition from dispersion-dominated to rotation-dominated kinematics in the ISM shown in Figure \ref{fig:vphi_sigma}) correlate with inner CGM virialization in both regimes, high-redshift disks are thicker and more turbulent.

    \item When galaxies are still in their bursty phase, accreted gas typically forms stars after residing in the galaxy for $\lesssim 5$ galaxy free-fall times (Figure \ref{fig:residencetimes_earlytime}).
    In contrast, in steady galaxies, gas is able to persist in the galaxy for up to $\sim25$ galaxy free-fall times before forming stars (Figure \ref{fig:residencetimes_latetime}).
    This highlights a key difference in how star formation is regulated in bursty galaxies, recently shown by \cite{sunTurbulentFrameworkStar2026} to be well modeled by a CGM-scale turbulent framework, as opposed to steady equilibrium disks.
\end{enumerate}

Our results in this work add to the growing body of evidence that gas accretion through the CGM may play a prominent role in galaxy formation and evolution.
As current and upcoming observatories, including JWST, Rubin, and Roman, increase our understanding of galaxy properties across cosmic time, a host of promising programs will shed new light on the properties of the CGM.
These include X-ray missions such as XRISM and NewAthena \citep{tashiroXRISMXrayImaging2022, barconsAthenaESAsXray2017}, cosmic microwave background experiments including Simons Observatory and CCAT Observatory \citep{adeSimonsObservatoryScience2019, ccat-primecollaborationCCATprimeCollaborationScience2023}, and fast radio burst projects such as CHIME/FRB, CHORD, and the Deep Synoptic Array \citep{chime/frbcollaborationCHIMEFastRadio2018, vanderlindeCanadianHydrogenObservatory2019, hallinanDSA2000RadioSurvey2019}.
A complete theoretical understanding of how the CGM couples to galaxy formation will be essential to interpret the observations.
In future work, it would be useful to derive predictions of observables from our simulations that could help probe the galaxy-CGM connection.
As our results investigate galaxy formation and CGM evolution using simulations with detailed stellar feedback models but without additional physics such as AGN feedback and cosmic rays, it would also be interesting to extend this analysis to simulations that include the additional physics.
This would be particularly relevant for massive ($\gtrsim L^*$) galaxies, for which AGN feedback likely plays a key role in quenching star formation rates, while the exact details on how the feedback couples to the CGM, and the effects of AGN feedback, magnetic fields, and cosmic rays on CGM-scale cooling flows, are still open questions.

\section*{Acknowledgements}
We thank the anonymous referee for their helpful comments, which improved the clarity of our manuscript.
I.S.\ was supported by the NSF Graduate Research Fellowship under Grant No.\ DGE-2234667.
C.-A.F.-G.\ was supported by NSF through grants AST-2108230 and AST-2307327; by NASA through grants 80NSSC22k0809, 80NSSC22K1124 and 80NSSC24K1224; by STScI through grant JWST-AR-03252.001-A; and by BSF through grant \#2024262.
J.S. was supported by the Israel Science Foundation (grant No. 2584/21), and by a grant from the US-Israel Binational Science Foundation \#2024262.
G.S. was supported by a CIERA Postdoctoral Fellowship.
This work was performed in part at the Aspen Center for Physics, which is supported by National Science Foundation grant PHY-2210452.
Numerical calculations were run on the Northwestern computer cluster Quest, the Caltech computer cluster Wheeler, Frontera allocation FTA-Hopkins/AST20016 supported by the NSF and TACC, XSEDE/ACCESS allocations ACI-1548562, TGAST140023, and TG-AST140064 also supported by the NSF and TACC, and NASA HEC allocations SMD-16-7561, SMD-17-1204, and SMD-16-7592.
Some of the calculations in this study utilize publicly available code developed by Alex Gurvich, which can be accessed at \url{github.com/agurvich/abg_python}.

\section*{Data Availability}
A public version of the GIZMO code is available at \url{http://www.tapir.caltech.edu/~phopkins/Site/GIZMO.html}. FIRE data products, including FIRE-2 simulation snapshots, initial conditions, and derived data products are available at \url{http://fire.northwestern.edu/data/}.

\bibliographystyle{mnras}
\bibliography{refs}

\appendix
\section{Temperature maps}\label{app:Tmaps}

\begin{figure*}
    \subfloat{\includegraphics[width=6.5in]{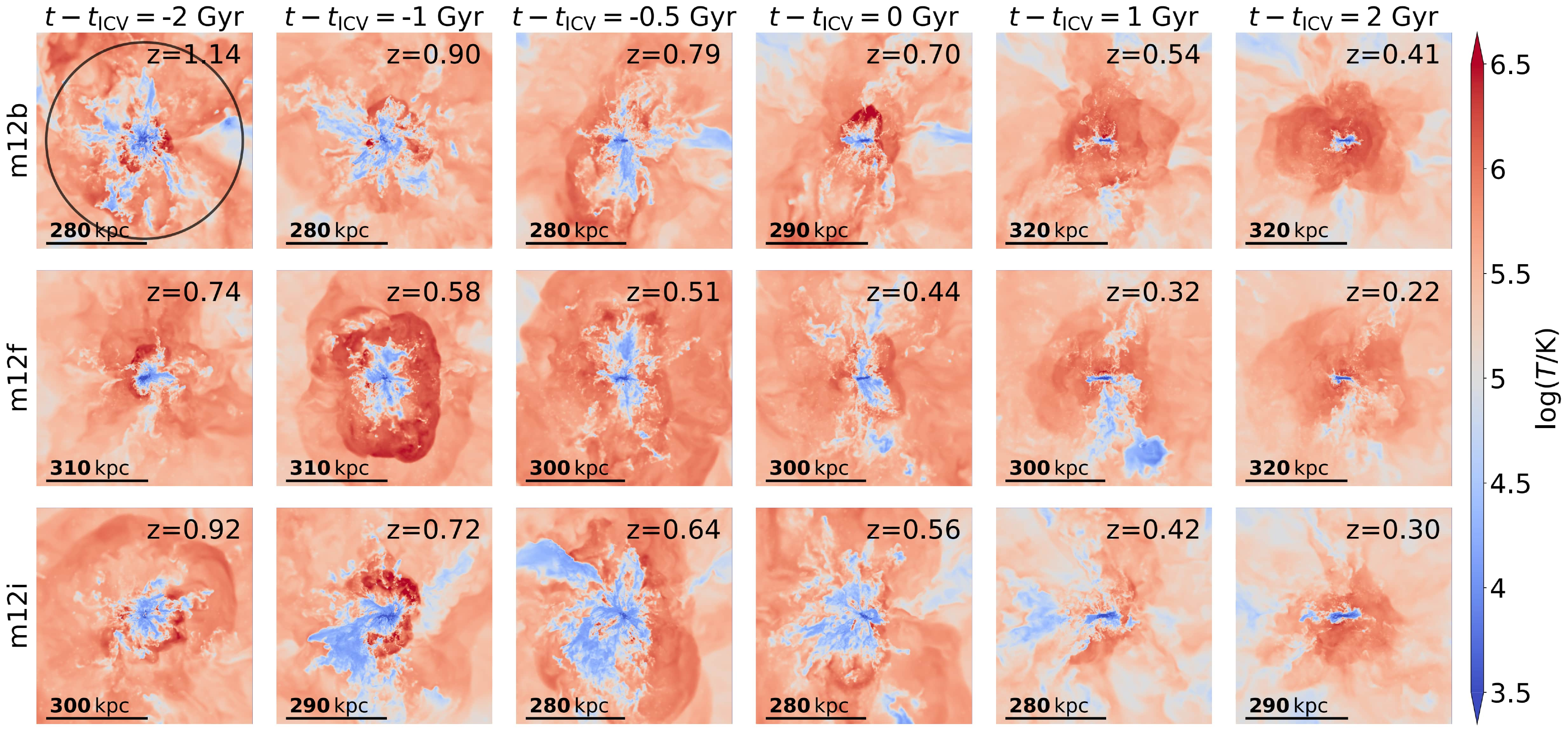}}
    \\[3em]
     \subfloat{\includegraphics[width=6.5in]{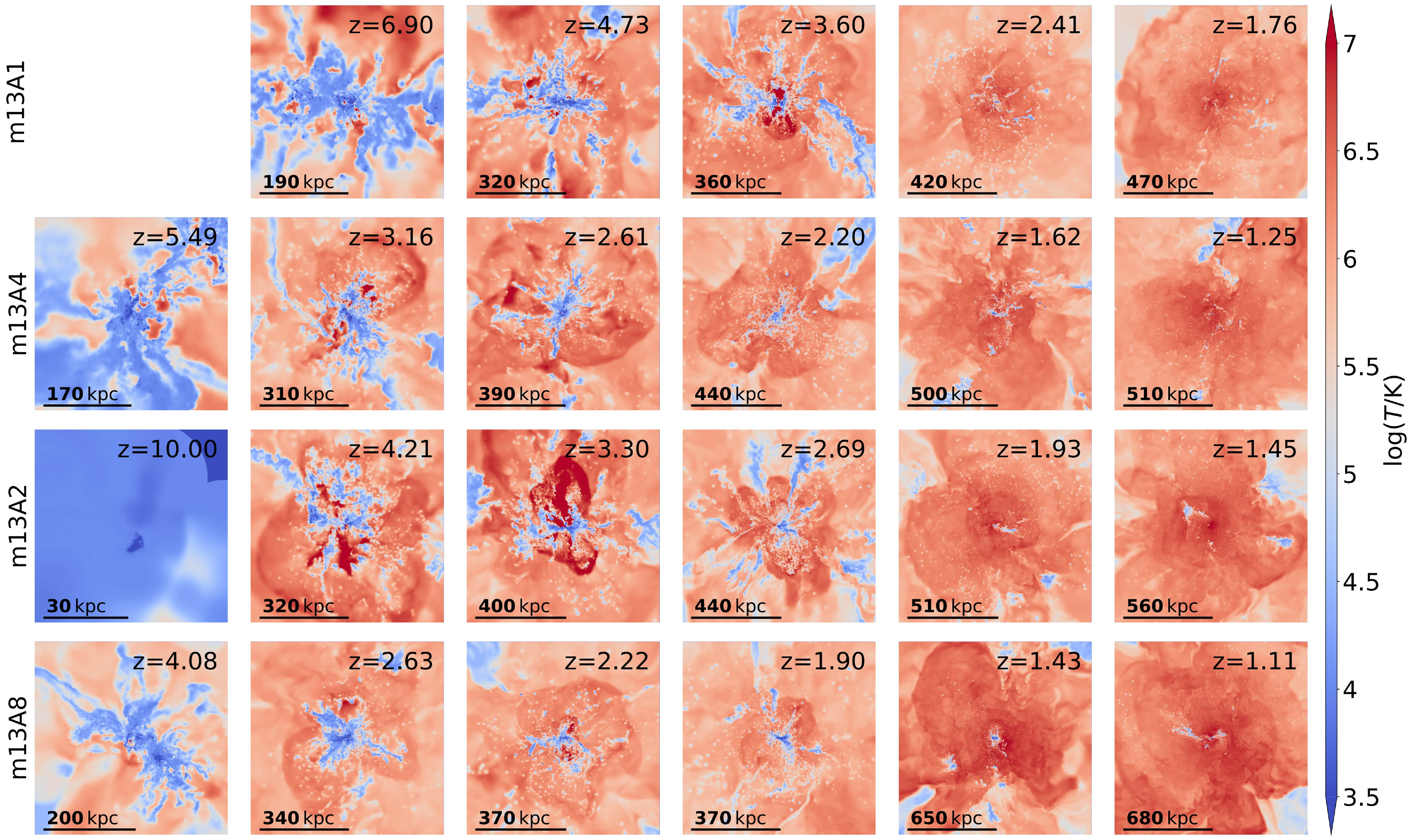}}
    \caption{Gas temperature maps for the seven halos in our analysis set.
    Projections of $\log T$ weighted by mass through a 30 comoving kpc slice perpendicular to the galactic plane are shown.
    Temperature maps at six snapshots are shown for each halo, spanning times 2 Gyr before ICV to 2 Gyr after ICV.
    Each panel has a side length of $2.2 R_{\mathrm{vir}}$, and the scale bar indicates $\sim R_{\mathrm{vir}}$ in units comoving kpc.
    The circle in the top left panel indicates the virial radius.
    The first panel for m13A1 is empty because the simulation starts at $t - t_{\mathrm{ICV}} \approx 1.8$.
    }
    \label{fig:Tmaps_halo}
\end{figure*}

\begin{figure*}
    \subfloat{\includegraphics[width=6.5in]{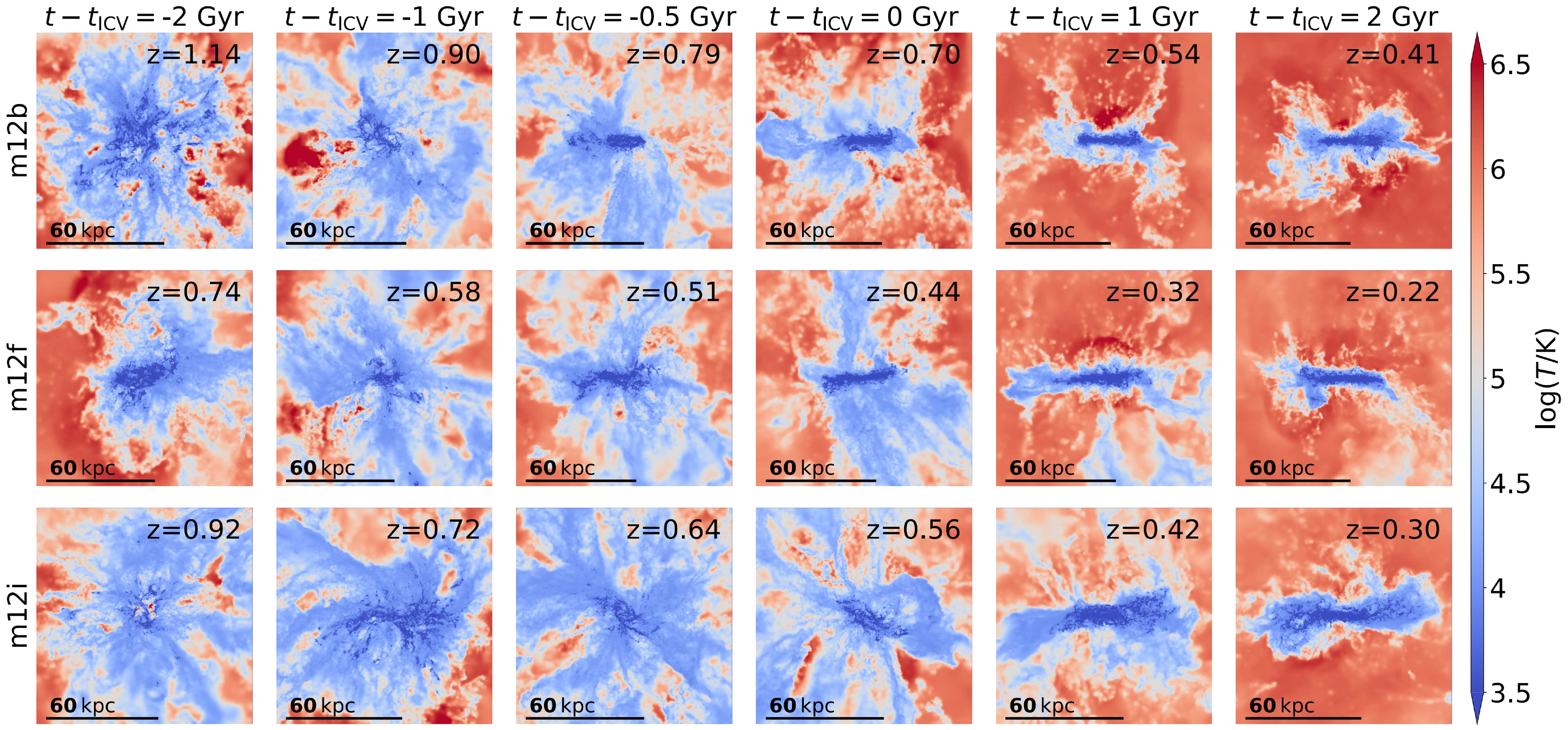}}
    \\[3em]
     \subfloat{\includegraphics[width=6.5in]{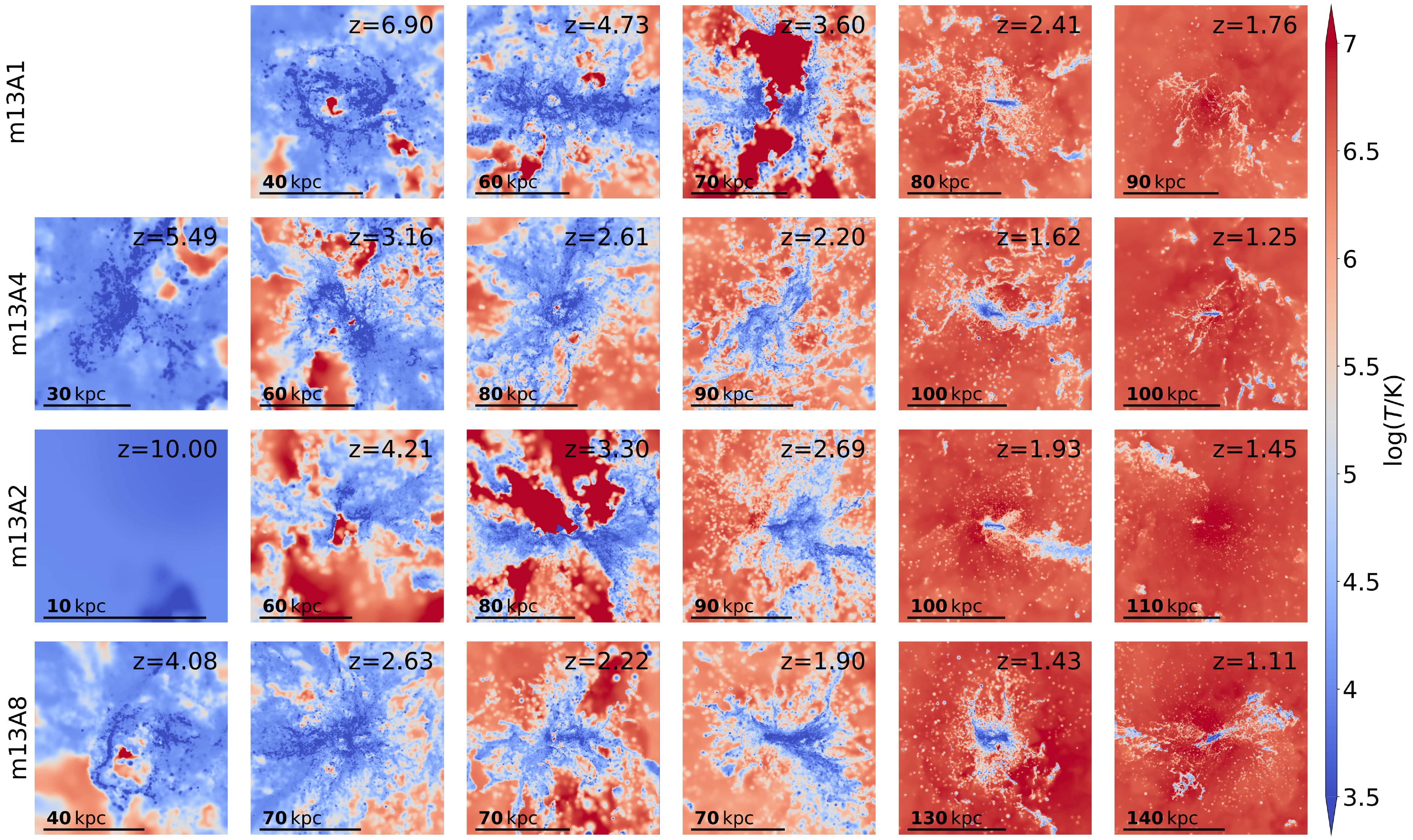}}
    \caption{Gas temperature maps of the inner CGM for the seven halos in our analysis set (see Table \ref{tab:sims}).
    Projections of $\log T$ weighted by mass through a 30 comoving kpc slice perpendicular to the galactic plane are shown, giving an edge-on view of each galaxy.
    The galactic plane is defined as the plane orthogonal to the rotation axis of the galaxy, which we compute from the total angular momentum of all star particles within $5 R_{s,1/2}$ (see Section \ref{sec:methods_halocentering}). 
    Temperature maps at six snapshots are shown for each halo, spanning times 2 Gyr before ICV to 2 Gyr after ICV.
    Each panel has a side length of $0.4 R_{\mathrm{vir}}$, and the scale bar indicates approximately $0.2 R_{\mathrm{vir}}$ in units comoving kpc.
    The halos initially contain significant amounts of cold gas in their inner regions ($\sim 0.1 R_{\mathrm{vir}}$).
    The inner halos become increasingly hot gas-dominated as the simulations are evolved.
    The first panel for m13A1 is empty because the simulation starts at $t - t_{\mathrm{ICV}} \approx 1.8$.
    }
    \label{fig:Tmaps}
\end{figure*}

In this section, we show maps of gas temperature for the full set of FIRE halos we analyze in this work.
Figure \ref{fig:Tmaps_halo} shows halo-scale maps (with panels of side length $2.2 R_{\mathrm{vir}}$), while Figure \ref{fig:Tmaps} shows a zoomed-in view of the inner halo (side length $0.4 R_{\mathrm{vir}}$).
The visualizations were produced with FIRE Studio \citep{gurvichFIREStudioMovie2022}, and we show maps at six snapshots that spam times 2 Gyr before ICV to 2 Gyr after ICV.
As in Figure \ref{fig:Tmaps_twoscales}, we show a mass-weighted projection of all gas inside a 30 comoving kpc slice perpendicular to the galactic plane.

Figure \ref{fig:Tmaps} shows that in the gigayears leading up to ICV, the inner CGM of the halos is dominated by cold $T\sim10^4$ K gas.
At around the time of ICV, the inner CGM becomes filled with hot, virial-temperature gas, while a cold disk (seen edge-on in the figure) forms in the center of the halo.

The post-ICV halos are not entirely devoid of cold gas even after the inner CGM virializes.
Figures \ref{fig:Tmaps_halo} and \ref{fig:Tmaps} show that at $t>t_\mathrm{ICV}$, the hot virialized halos contain some cold gas.
The maps depict a messy, complex cold gas structure in our simulations following ICV.
This differs from some analytic models of massive halos at high redshift, in which galaxies are fed by a small number of neat and ordered cold streams.

\section{ICV and the Bursty-to-Steady transition}\label{app:tburstytICV}
\begin{figure}
	\includegraphics[width=3.12in]{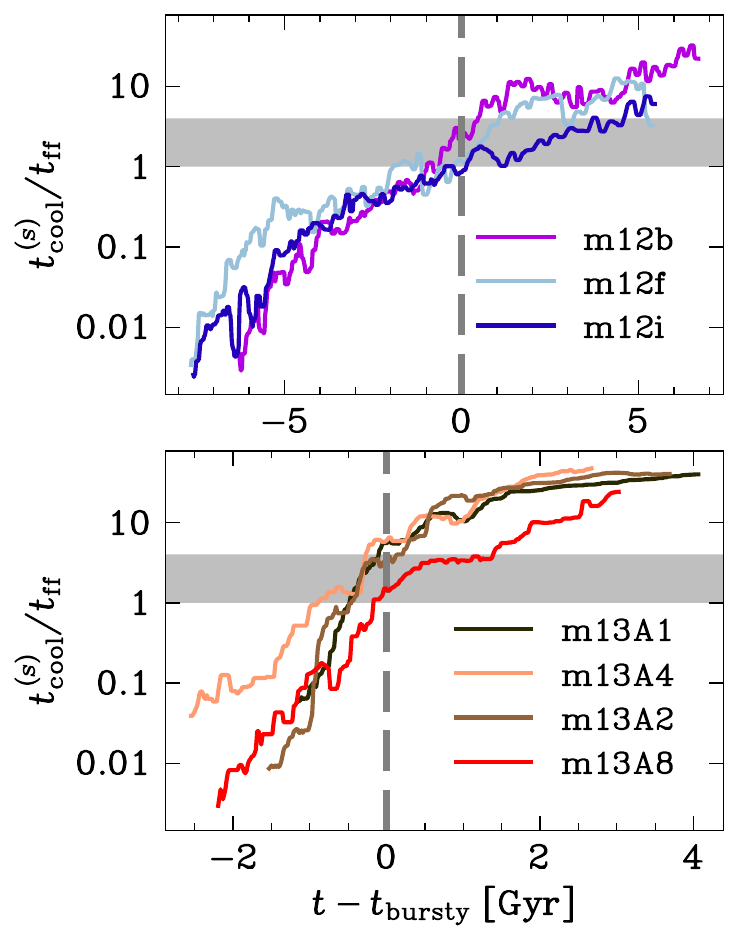}
    \caption{Time evolution of the ratio of the cooling time of shocked gas to the free fall time at $0.1 R_\mathrm{vir}$.
    The upper panel shows the low-redshift MW-mass halos we analyze, while the lower panel shows the high-redshift more massive halos.
    The results are plotted as a function of cosmic time relative to $t_\mathrm{bursty}$, the time when the SFR becomes steady.
    The shaded region indicates $1 \le t_\mathrm{cool}^{(s)}/t_\mathrm{ff} \le 4$.
    At these ratios, the inner CGM is expected to virialize.
    Generally, the inner CGM virializes at roughly the same time that star formation in the galaxy transitions from being bursty to steady.
    }
    \label{fig:tbursty_tICV}
\end{figure}

In this section, we test how well $t_\mathrm{bursty}$ correlates with the time of virialization of the inner CGM.
In Figure \ref{fig:tbursty_tICV}, we show $t_\mathrm{cool}^{(s)}/t_\mathrm{ff}$ measured at $0.1R_\mathrm{vir}$ as a function of $t-t_\mathrm{bursty}$, as in \cite{gurvich_rapid_2023}.
As described in Section \ref{sec:tICV}, $t_\mathrm{cool}^{(s)}/t_\mathrm{ff}$ is an indicator of the time when the inner CGM virializes.
Specifically, \cite{sternVirializationInnerCGM2021} found that the inner CGM is expected to virialize when $1 \le t_\mathrm{cool}^{(s)}/t_\mathrm{ff} \le 4$, which is represented by the shaded band in Figure \ref{fig:tbursty_tICV}.

We find that $t_\mathrm{bursty}$ and $t_\mathrm{ICV}$ occur at roughly the same time.
In Figure \ref{fig:tbursty_tICV}, $t_\mathrm{cool}^{(s)}/t_\mathrm{ff}$ generally crosses the shaded band at around $t_\mathrm{bursty}$.

\section{Radial profiles of mass inflow rates}\label{app:Mdotprofiles}
\begin{figure*}
	\includegraphics[width=6.5in]{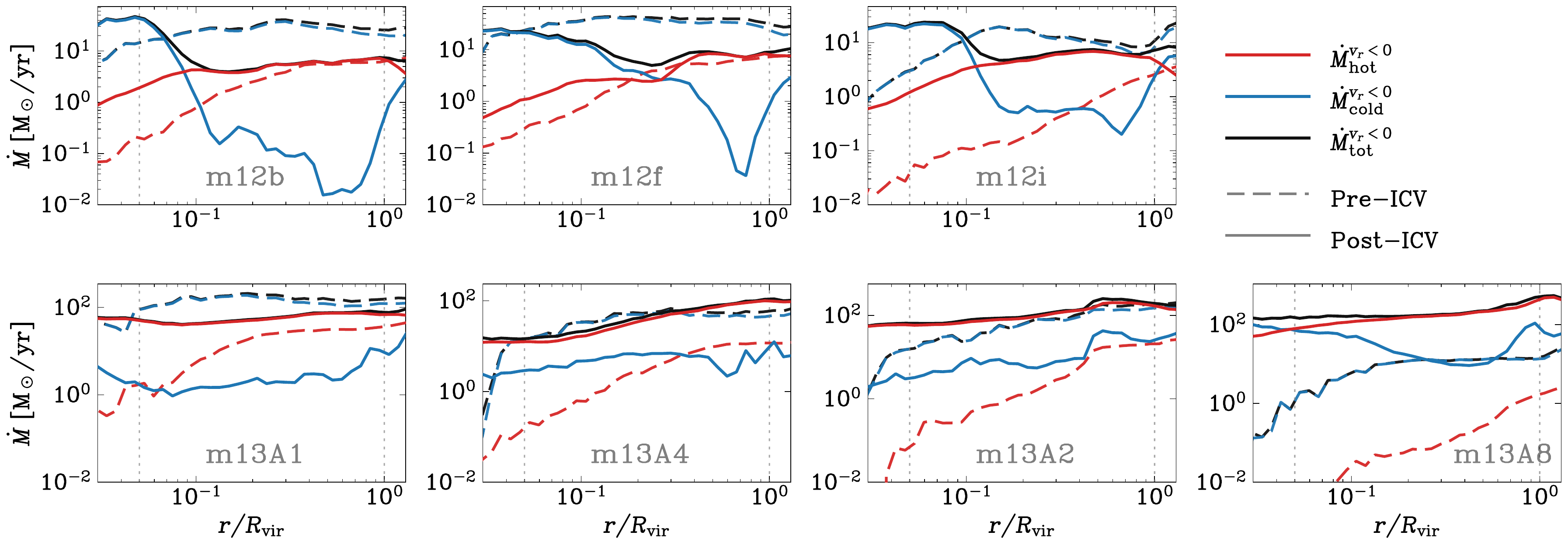}
    \caption{Radial profiles of mass inflow rates in the CGM.
    For the seven halos in our analysis set, mass flow rates of inflowing gas ($v_r<0$) are shown for cold ($T \leq 10^5$~K; blue curves), hot ($T>10^5$~K; red curves), and all (black curves) gas.
    The dashed lines indicate the median profiles averaged in a 400 Myr time interval before ICV ($2.4 \lesssim z \lesssim 2.7$ for the m12 halos, and $5.7 \lesssim z \lesssim 4.4$ for the m13 halos).
    The solid lines indicate the median profiles averaged in a 400 Myr time interval after ICV ($0 \lesssim z \lesssim 0.03$ for the m12 halos, and $1 \lesssim z \lesssim 1.1$ for the m13 halos).
    The vertical lines mark $0.05 R_\mathrm{vir}$ and $R_\mathrm{vir}$.
    In pre-virialized halos, cold gas dominates the inflow, whereas inflows in the CGM ($r \gtrsim 0.1 R_\mathrm{vir}$) are dominated by hot gas in post-ICV halos.
    The mass flow rate of hot infalling gas increases with radius on average in both pre- and post-ICV halos.
    }
    \label{fig:Mdotprofiles}
\end{figure*}
In this section, we investigate the radial dependence of mass flow rates of CGM inflows in our simulations.
This builds on the results shown in Figure \ref{fig:Mdotin_allz}, in which we presented mass flow rates in three discrete radial bins extending from $0.1 R_\mathrm{vir}$ to $0.6 R_\mathrm{vir}$.
Figure \ref{fig:Mdotprofiles} shows radial profiles of inflowing gas in the CGM, for a pre-ICV time interval and a post-ICV time interval.
The black curves show the total mass flow rates of all inflowing gas, while the red and blue curves show results for hot and cold gas, respectively. 
This figure reinforces our result that cold gas dominates the inflow in pre-virialzied halos, while most of the inflow in the CGM ($r \gtrsim 0.1 R_\mathrm{vir}$) is hot in post-ICV halos.
Additionally, on average, the mass flow rate of hot infalling gas increases with radius for both the pre- and post-ICV time intervals shown.
On the other hand, the flow rate for all infalling gas (dominated by cold gas) in pre-ICV halos tends to be much more stable with radius in the CGM.
Thus, a radial gradient in $\dot{M}_{\mathrm{hot}}^{v_r<0}$ primarily drives the radial dependence of $\left(\dot{M}_{\mathrm{hot}}/\dot{M}_{\mathrm{tot}}\right)^{v_r<0}$ at redshifts before ICV that is found in Figure \ref{fig:Mdotin_allz}, where hot fractions are systematically larger in the larger radii bins than the smaller radii bins.

\section{Star formation times}\label{app:tformtICV}
\begin{figure*}
	\includegraphics[width=6.5in]{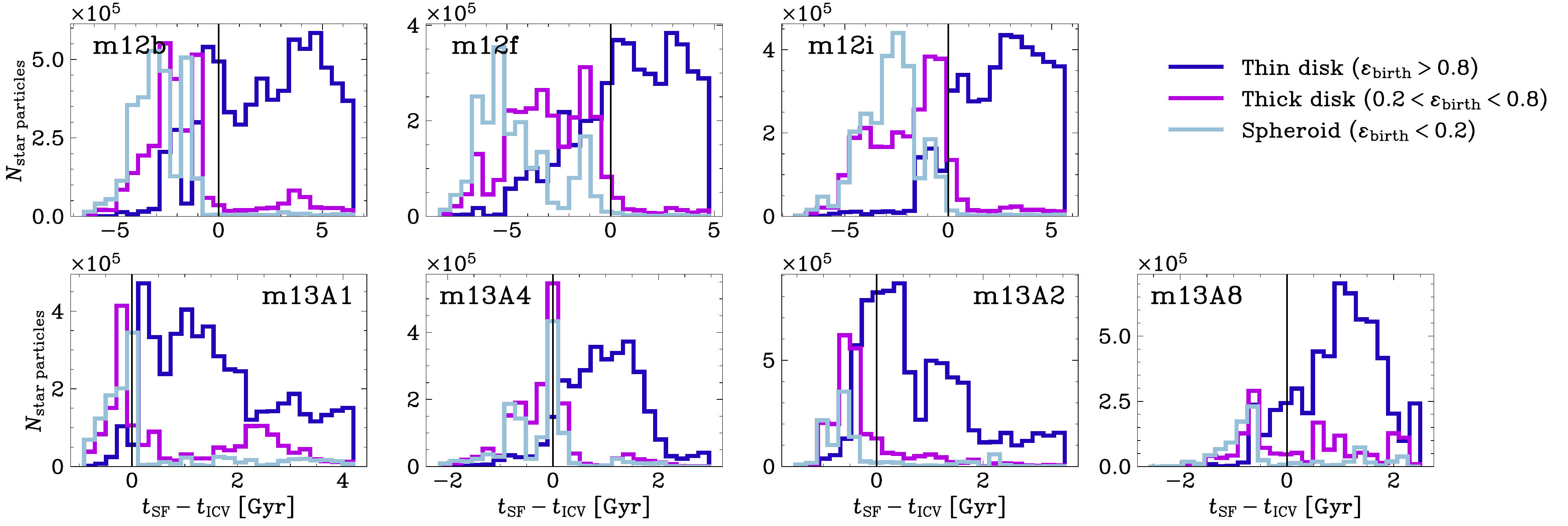}
    \caption{Distributions of the formation times of stars in the central galaxy at the final simulation snapshot, relative to the time of ICV of their halo.
    The star particles are separated into spheroid ($\epsilon_\mathrm{birth}<0.2$), thick disk ($0.2<\epsilon_\mathrm{birth}<0.8$), and thin disk ($\epsilon_\mathrm{birth}>0.8$) classes based on their circularity at birth, $\epsilon_\mathrm{birth}$.
    Only \textit{in situ} stars (i.e., stars that formed within $0.1R_\mathrm{vir}$ of the most massive progenitor at their formation time) are included.
    Spheroid and thick disk star formation peaks before ICV, while thin disk stars are predominantly formed after ICV.
    }
    \label{fig:starformationtimes}
\end{figure*}

We investigate when stars belonging to spheroid, thick disk, and thin disk populations form with respect to the time of virialization of their inner halo.
As we detail in Section \ref{sec:results_startracking}, we assign \textit{in situ} stars (i.e., stars that formed within $0.1R_\mathrm{vir}$ of the most massive progenitor at their formation time) to the three classes based on their circularity at birth, using the criteria of \cite{yuBornThisWay2023}.

In Figure \ref{fig:starformationtimes}, we show distributions of the formation times of spheroid, thick disk, and thin disk stars.
Generally, spheroid and thick disk star formation peaks before ICV, and subsides after ICV.
Thin disk stars dominate star formation after ICV.
Our results are consistent with previous studies by, e.g., \cite{yuBornThisWay2023}, who found a similar trend in MW-mass FIRE galaxies when star formation is measured with respect to $t_\mathrm{bursty}$, which we find is correlated with $t_\mathrm{ICV}$ (see Appendix \ref{app:tburstytICV}).

\end{document}